%% file: doc.tex
\PassOptionsToPackage{dvipsnames,svgnames}{xcolor}

\documentclass[acmsmall,screen,nonacm]{acmart}

\usepackage{subcaption}
\usepackage{float}
\usepackage{tabulary}
\usepackage{threeparttable}
\usepackage{booktabs}
\usepackage{relsize,xspace}
\usepackage{multirow}
\usepackage{enumitem}
\usepackage{balance}
\usepackage{array}
\usepackage{ltablex}
\usepackage{hyperref}
\usepackage{multicol}
\usepackage{microtype}
\usepackage{makecell}
\usepackage{graphicx}
\usepackage{pgfplots}
\usepackage[capitalize]{cleveref}
\crefname{section}{Sec.}{sections}
\Crefname{section}{Section}{Sections}

\usepackage[textsize=tiny]{todonotes}

\newcommand\parhead[1]{\vspace{.5mm}\noindent\textbf{{#1}}}

\begin{document}

\newcommand{\tb}[1]{{\textsf{\textbf{TB}}[\smaller\sffamily\color{orange} #1]}}
\newcommand{\sven}[1]{{\textsf{\textbf{SP}}[\smaller\sffamily\color{purple} #1]}}
\newcommand{\kh}[1]{{\textsf{\textbf{KH}}[\smaller\sffamily\color{blue} #1]}}
\newcommand{\rwi}[1]{{\textsf{\textbf{RW1}}[\smaller\sffamily\color{red} #1]}}
\newcommand{\rwii}[1]{{\textsf{\textbf{RW2}}[\smaller\sffamily\color{red} #1]}}
\newcommand{\rwiii}[1]{{\textsf{\textbf{RW3}}[\smaller\sffamily\color{red} #1]}}

\newcommand{\quotebox}[3]{\vspace{.5em}\noindent\begin{tikzpicture}
    \node[align=center,draw,thin,minimum width=\columnwidth,inner sep=2.1mm] (titlebox)%
    {\parbox{0.98\columnwidth}{\looseness=-1\noindent\textit{#2\vspace{0.1cm}}}};
    \node[label=left:{\colorbox{white}{\small #1}}] (W) at (titlebox.south east) {};%
    \end{tikzpicture}\vspace{-18pt}}

\newcommand{\sumbox}[1]{
\noindent\fbox{\parbox{\dimexpr \linewidth-2\fboxsep-2\fboxrule}{\noindent
		\textit{#1}
	}}\smallskip
}

\title{Many Tools, Few Exploitable Vulnerabilities: A Survey of 246 Static Code Analyzers for Security}

\acmJournal{CSUR}

\author{Kevin Hermann}
\orcid{0009-0004-6207-4045}
\affiliation{%
  \institution{Ruhr-University Bochum}
  \country{Germany}
}
\email{kevin.hermann@rub.de}

\author{Sven Peldszus}
\orcid{0000-0002-2604-0487}
\affiliation{%
  \institution{Chalmers\,|\,University of Gothenburg}
  \country{Sweden}
}
\email{sven.peldszus@gu.se}

\author{Thorsten Berger}
\orcid{0000-0002-3870-5167}
\affiliation{%
  \institution{Ruhr-University Bochum}
  \country{Germany}
}
\affiliation{%
  \institution{Chalmers\,|\,University of Gothenburg}
  \country{Sweden}
}
\email{thorsten.berger@rub.de}



\begin{abstract}\looseness=-1
Static security analysis is a widely used technique for detecting software vulnerabilities across a wide range of weaknesses, application domains, and programming languages.
While prior work surveyed static analyzes for specific weaknesses or application domains, no overview of the entire security landscape exists.
We present a systematic literature review of 246 static security analyzers concerning their targeted vulnerabilities, application domains, analysis techniques, evaluation methods, and limitations.
We observe that most analyzers focus on a limited set of weaknesses, that the vulnerabilities they detect are rarely exploitable, and that evaluations use custom benchmarks that are too small to enable robust assessment.
\end{abstract}

\maketitle

\input{01_Introduction.tex}

\input{02_Background}

\input{related} 

\input{03_Methodology.tex}
\input{06_Limitations.tex}

\input{04_Results.tex}

\input{05_Discussion.tex}

\input{07_Conclusion.tex}


\bibliographystyle{IEEEtran}
\bibliography{IEEEabrv,references}

\end{document}

%% file: 01_introduction.tex
\section{Introduction}
Software security has become a primary concern for modern systems\,\cite{hermann2025exploratory},
as vulnerabilities increasingly threaten their reliability, integrity, and operation\,\cite{bau2012, egele2013, lazar2014, nadi2016, fahl2013, krombholz2017, roth2021}.
To mitigate these risks, security must be systematically engineered into software and continuously validated through rigorous vulnerability assessments\,\cite{McGraw04_software_security}.
A key enabler of this process is static security analysis, which automates the detection of potential vulnerabilities during development\,\cite{Potter2004,black2012}.
In fact, static analysis is the most popular technique for validating security properties as part of secure software engineering\,\cite{Khan2021}.
Consequently, both researchers and practitioners have presented a huge number of static security analyzers that address numerous types of weaknesses.

To systematically address these weaknesses, frameworks such as the Common Weakness Enumeration (CWE) provide a standardized taxonomy for understanding, categorizing, and prioritizing security vulnerabilities.\,\cite{cwe_mitre}.
Similarly, initiatives like the  OWASP Top 10\,\cite{owasp_top_ten} identify the most critical and prevalent security risks in web applications, offering practical guidance for developers and security practitioners. 
These frameworks do not only serve as essential references for identifying and prioritizing security flaws, but also inform the design of static security analyzers, many of which explicitly map their detection capabilities to specific CWEs.
However, while these initiatives provide a clear overview of existing vulnerabilities and weaknesses, a critical gap remains: \emph{we lack a comprehensive understanding of which vulnerabilities can be automatically detected by static analyzers and which require manual intervention}.

Existing surveys have explored static security analyzers and related techniques. However, they are limited to specific domains, such as Android\,\cite{zhu2024} or IoT applications\,\cite{Feng2023,Xie2017}, to specific CWEs\,\cite{Adhikari2025}, or to papers published in certain security venues\,\cite{cui2022}.
To the best of our knowledge, \emph{there exists no comprehensive overview of static security analyzers across the whole security domain}.

To bridge this gap and provide developers with actionable guidance, we must systematically examine which static analyzers are available, what vulnerabilities they can detect, how they operate, and where their limitations lie. Such an analysis will not only clarify the current capabilities of automated detection, but also highlight critical areas for future research and tool development.

Our survey addresses this gap with a large-scale systematic literature review of static security analysis techniques, assessing the state of the art. It answers the following research questions:
\vspace{-1pt}
\begin{enumerate}
    \item[RQ1:] \emph{What static security analyzers have been presented in the literature and what domains do they address?} While some overviews exist, they are often limited to specific application domains or vulnerability types. We provide a comprehensive review across the whole security domain.
    \item[RQ2:] \emph{What vulnerabilities do static security analyzers address?} Analyzers are often tailored to specific vulnerabilities, e.g., injection flaws or memory safety issues. We investigate which vulnerabilities are most commonly or less frequently addressed.
    \item[RQ3:] \emph{How do static security analyzers scan code for vulnerabilities?} Analyzers can use various techniques to analyze code, e.g., dataflow analysis or pattern matching. We investigate the techniques and what kind of intermediate representations are used by them.
    \item[RQ4:] \emph{How are static security analyzers evaluated?} Analyzers can be evaluated via benchmarks, real-world case studies, and many other ways. We determine current evaluation practices.
    \item[RQ5:] \emph{What are the limitations of static security analyzers?} Analyzers may be limited in scope, may produce false positives, among many other limitations. We elicit the reported limitations.
\end{enumerate}
\vspace{-1pt}
Our survey covers 246 static security analyzers, in total capable of detecting 161 CWEs, and their evaluation in 347 empirical evaluations.
We learned that most analyzers are domain-independent and focus on a small set of well-known vulnerabilities, such as injection flaws and memory safety issues, leaving many others understudied.
Evaluations solely rely on small, custom benchmarks that limit generalizability, and many tools fail to report their limitations altogether.

Our survey can be used by researchers to investigate open research gaps in static security analysis.
Practitioners can use our results to identify existing static security analyzers that fit their needs.
We describe concepts and techniques used in static security analysis, which can help practitioners to better understand how static security analyzers work.

%% file: 02_Background.tex
\section{Background and Related Work}  

We now present the relevant background and related work.

\subsection{Application Security and Vulnerabilities}\looseness=-1
Almost all software systems need to consider security nowadays.
However, developers are typically not security experts\,\cite{Green2016} and often prioritize functionality and time-to-market over security concerns\,\cite{Naiakshina2017,hermann2025exploratory}. 
As a result, software systems frequently contain exploitable vulnerabilities.

A  \emph{vulnerability} is a flaw in a software system that can be exploited by attackers\,\cite{ISO27000,SP800-28} to compromise the confidentiality, integrity, or availability of the system or its data\,\cite{Elder2024}.
While a vulnerability refers to a specific instance, a \emph{weakness} denotes an underlying kind of defect in design or implementation of which the vulnerability is an instance. 
Not all software bugs constitute security vulnerabilities, but many vulnerabilities originate from general programming defects\,\cite{Bojanova2023}.

To systematically document and classify vulnerabilities, the security community has established several widely adopted standards and knowledge bases. 
While the Common Vulnerabilities and Exposures (CVE)\,\cite{cve} is a list of security vulnerabilities, the Common Weakness Enumeration (CWE)\,\cite{cwe_mitre} is a taxonomy of recurring software weaknesses that can lead to vulnerabilities.
In the web domain, the OWASP\,\cite{owasp_top_ten} maintains a list of the top 10 most critical web application security risks, which is a widely used reference for web application security.
These catalogs help developers and security professionals understand and mitigate common security issues in software systems.

Vulnerabilities may arise from incorrect implementation or misuse of dedicated security features, such as access control mechanisms or cryptographic libraries. 
Security features are functionalities that protect a system from malicious attacks or protect sensitive data\,\cite{hermann2025taxonomy}.
The mere presence of security features does not guarantee security, but improper configuration, integration, or usage can itself introduce exploitable flaws\,\cite{Acar2017}. 
More broadly, vulnerabilities often stem from insecure implementation practices, including insufficient input validation, memory mismanagement, race conditions, or logic errors. 
Classical examples include buffer overflows caused by improper memory handling, as well as injection flaws resulting from inadequate sanitization of untrusted input.

In addition to first-party code, modern software systems heavily rely on third-party dependencies. 
Particularly, security features are implemented using specific libraries\,\cite{hermann2025exploratory}.
Consequently, vulnerabilities may also originate from external dependencies or be introduced through supply-chains\,\cite{Williams2025}. 

Overall, vulnerabilities emerge from the interplay between developer practices, system complexity, and insufficient integration of security considerations into software implementations. 
Therefore, beyond implementing security features, secure development processes, systematic testing, and continuous vulnerability management are essential to ensure secure software systems.

\subsection{Vulnerability Detection Techniques}
To mitigate security vulnerabilities, various techniques and tools have been developed to detect and fix vulnerabilities in software systems.
These techniques often include dynamic application security testing, static application security testing, or machine-learning-based techniques.

\textbf{Dynamic application security testing} techniques analyze a software system during its execution to identify security vulnerabilities.
For example, Deemon\,\cite{Pellegrino2017} is a dynamic analysis framework for web applications that models application behavior using property graphs and identifies vulnerabilities by searching for specific graph patterns.

\textbf{Fuzzing}\,\cite{sutton2007,Takanen2018} has become a popular technique to find security vulnerabilities in software systems.
Blackbox-fuzzing\,\cite{Takanen2018} generates inputs randomly based on predefined rules without any knowledge of the program internals.
In contrast, whitebox-fuzzing\,\cite{Takanen2018} treats inputs as symbolic values and uses constraint solvers to explore different program paths.
However, fuzzing may not cover all code paths, and miss vulnerabilities that require specific input conditions\,\cite{Ognawala2018}.

\textbf{Model-based security testing} (MBST) techniques\,\cite{Ramakrishnan2002,PariSalas2007,Lebeau2013,felderer2016} use abstract models of a system to generate test cases that can be executed dynamically to find security vulnerabilities.
Creation and maintenance of accurate models can be time-consuming and error-prone in practice\,\cite{felderer2016}.

\textbf{Penetration testing}\,\cite{arkin2005,Potter2004} is the manual or semi-automated simulation of real-world attacks to identify vulnerabilities. 
It is time-consuming and costly\,\cite{hermann2025exploratory}, and its effectiveness depends heavily on the testers' expertise\,\cite{Potter2004}.
While penetration testing can reveal complex, exploit-driven vulnerabilities, it may overlook even simple issues that automated tools can detect systematically.

\textbf{Runtime instrumentation and monitoring} techniques are dynamic analyses that detect security errors during program execution. 
They instrument code to track additional metadata (e.g., shadow values) for detecting memory errors\,\cite{Nethercote2007} or insert runtime security checks\,\cite{Peldszus2024}. 
Similarly, sanitizers are dynamic analysis tools that instrument programs to detect various types of errors at runtime, such as memory errors, undefined behavior\,\cite{Song2019}.
However, due to their heavyweight instrumentation, these techniques typically incur substantial runtime overhead\,\cite{Peldszus2024}.

\textbf{Machine learning based techniques} treat source code as text or graph structures and use machine learning models, e.g., neural networks\,\cite{Lin2020} or large language models\,\cite{Zhou2025}, to classify code snippets as vulnerable or non-vulnerable\,\cite{Shiri2024}.
However, since these techniques rely on patterns learned from training data, they often struggle generalize beyond benchmark datasets\,\cite{Risse2024}.
Due to lacking quality of underlying datasets, their effectiveness is often worse than random guessing on real systems\,\cite{Ding2024}.
Recent studies further show that large language models may appear to detect vulnerabilities even when the code itself provides insufficient semantic evidence, suggesting spurious correlations in the learned representations\,\cite{Risse2025}.

\textbf{Static application security testing} (SAST) techniques, such as static code analysis, code reviews, or code auditing, analyze source code, bytecode, or binaries without executing the program to identify potential security vulnerabilities\,\cite{Potter2004}.
In this paper, we focus on static analysis techniques, which are automated tools that analyze source code to identify potential security vulnerabilities and that can be used early in the software development lifecycle\,\cite{black2012}.
Their capabilities and their usability have been studied in a range of studies\,\cite{AlShammare2025}.
In fact, practitioners require comparisons between SAST tools on benchmarks, as well as customizations of these tools\,\cite{Li2025}.
Typically, developers must configure tools based on the project at hand, and prioritize warnings from SAST tools based on the development context\,\cite{vassallo2020developers,piskachev2023}.
However, developers' experience and security expertise can also influence their ability to effectively identify vulnerabilities from analysis reports\,\cite{baca2009}.

\subsection{Static Security Analysis}
Static security analysis examines source code without executing it to identify potential vulnerabilities. 
These tools can be applied early in the software development lifecycle, allowing developers to detect and remediate issues before deployment\,\cite{black2012}.
Various techniques exist, ranging from pattern- or rule-based checks for insecure coding practices, to more advanced approaches such as data-flow analysis, control-flow analysis, symbolic execution, or abstract interpretation, which explore potential program paths to uncover deeper vulnerabilities\,\cite{Benjamin2005Finding,Chess2004}. 

Static analyzers can operate on various artifacts, including source code, bytecode, or configuration files, making them applicable to a broad range of systems\,\cite{Benjamin2005Finding,Diaz2013}. 
They target many types of security flaws, such as memory errors (e.g., buffer overflows), input validation failures leading to injection attacks, authentication or access-control mistakes, and cryptographic misconfigurations\,\cite{Pistoia2007,Gomes2025}. 
However, a comprehensive overview of covered weaknesses and what can be analyzed is missing.

To effectively detect security vulnerabilities using static analysis, developers should choose a suitable combination of static analysis tools based on their specific project requirements and context\,\cite{nunes2019}.
For example, a check may look for the use of insecure functions, or whether user input is properly sanitized before being used in a database query.
Although static analysis techniques can be effective in identifying potential security vulnerabilities early, they may produce a large number of false positives\,\cite{Peldszus2026}, or alerts which are hard to understand for developers\,\cite{vassallo2020developers}.

%% file: related.tex
\subsection{Related Work}
Dalaq et al.\,\cite{Dalaq2025} study benchmarks and evaluation metrics of 57 SAST tools.
We confirm their findings on metrics and reusable benchmarks for evaluating SASTs and additionally investigate sample sizes for benchmarks, non-reused benchmarks, and case studies.
Cui et al.\,\cite{cui2022} survey various vulnerability detection methods, including static analysis, published at security conferences.
Our study also includes static analysis tools published at further conferences and journals of computer science, such as the software engineering domain.
Multiple works\,\cite{Xie2017,Feng2023,zhu2024,Gomes2025} survey vulnerability detection methods for the IoT domain.
In addition, Zhu et al.\,\cite{zhu2024} investigate the effectiveness of SAST tools for Android on a unified benchmark.
Our survey extends these findings by including domain-independent tools, as well as those for other domains.
Adhikari et al.\,\cite{Adhikari2025} investigate the effectiveness of static analysis tools in detecting CWEs related to memory vulnerabilities, improper input validation, and hardcoded credentials.
Although we do not evaluate the effectiveness of tools, we provide a list of tools that are capable of detecting each CWE, and organized them according to the CWE comprehensive categorization.
Jerónimo et al.\,\cite{Jeronimo2024} perform a literature survey of 18 static analyzers usable for early vulnerability detection, presenting their techniques, and limitations.
Our findings align with theirs, but extend them with further techniques.

%% file: 03_Methodology.tex
\section{Methodology}
We conducted a systematic literature review by following the ACM SIGSOFT's Empirical Standards\,\cite{acm_standards}. 
We describe how we selected and analyzed static analysis techniques in the following.

\subsection{Static Analyzer Selection and Data Extraction}
To answer our research questions, we searched for scientific publications that present static analysis techniques for vulnerability detection in code.
During our selection process, we applied inclusion criteria (IC) and exclusion criteria (EC) to decide whether to include a paper for our study.

\parhead{Inclusion Criteria:}
To study the landscape of static security analysis, we included papers that:
\begin{itemize}[leftmargin=26pt]
	\item[\textit{IC1:}] contribute a novel static analysis or a technique related to vulnerability detection in code, or
	\item[\textit{IC2:}] an application or extension of a static analysis technique for vulnerability detection in code.
\end{itemize}

\parhead{Exclusion Criteria:}
To ensure relevance to static security analysis, we excluded papers that:
\begin{itemize}[leftmargin=29pt]
	\item[\textit{EC1:}] do not focus on a static analysis technique for vulnerability detection,
	\item[\textit{EC2:}] only present surveys or reviews related to static security analyzers, or
	\item[\textit{EC3:}] present only theoretical concepts without an executable technique.
\end{itemize}

To identify relevant papers, we queried the digital library Scopus using the following search string: ``static AND security AND analysis''.
We performed the search on the title and abstract of papers published until December 31, 2024.
To focus on peer-reviewed research, we limited our search to conference papers and journal articles written in English and categorized under the field of computer science.
This search yielded a total of 6,661 papers.

We then applied a four-step data extraction process, containing filtering based on title, abstract, and full paper to identify static analysis techniques for detecting vulnerabilities in code.
We used this process to align the authors' understanding by independently filtering and discussing discrepancies.
We report agreement scores to show the effect of this alignment, but did not conduct independent coding of disjunct samples.
All discrepancies have been discussed among the authors.

\textbf{Step 1:} The first and second authors manually inspected the titles of the 6,661 papers identified in the search, independently rating whether to include each paper.
All conflicting decisions were discussed and resolved collaboratively, but in ambiguous cases, the paper was included for further investigation in the next phase.
To facilitate frequent discussions, the papers were inspected in multiple batches.
This process aimed to establish a shared understanding among the raters and ensure consistency in the coding.
The inter-coder agreement in the first batches was 88.7\,\% by percent agreement, with a Cohen’s Kappa of 0.558, indicating moderate reliability.
In the final batch of 447 papers, the authors reached substantial agreement, with 92.8\,\% agreement and a Cohen’s Kappa of 0.675.
The filtering resulted in 920 papers considered in the second filtering phase.

\textbf{Step 2:} We then applied the inclusion and exclusion criteria to the abstracts of the remaining papers.
We started this filtering step with a batch of 50 papers, whose abstracts were independently read by the first and second author, reaching an agreement of 86\,\% and a Cohen's Kappa of 0.685, indicating substantial agreement.
Still the discussions about conflicts aided in further clarifying the criteria for including papers, e.g., we excluded ML-based analyses that are only data-driven.
Thereafter, the first author continued the filtering and only problematic cases were discussed.
To validate the alignment between the authors, at the end of the abstract-based filtering, we drew a second sample of 50 papers that were independently inspected by the second author.
On this sample, we observed an agreement of 98\,\% and an Cohen's Kappa of 0.96, showing almost perfect agreement.
Overall, this resulted in 422 papers considered for filtering based on the full paper.

\textbf{Step 3:} Due to the observed almost perfect agreement, the first author read the full papers actively involving the second author via discussions of non-obvious papers.
In this step, we excluded papers out of scope based on the full content of the paper (156 papers), or were not available through our institution (20 papers).
This filtering step resulted in 246 papers for the final review.


\begin{table}[t]
	\caption{Categories and example codes used for open coding of static security analyzers}
	\label{tab:coding_categories}
	\vspace{-.4cm}
	\scriptsize
	\begin{tabular}{ll}
		\toprule
		\textbf{Category}           & \textbf{Example Codes}                                                                                         \\
			\midrule
		Name                        & \textless{}name of the technique\textgreater{}                                                          \\
		Domain                      & domain-independent, web applications, android, operating systems, iot, \dots                              \\
		Language                    & c/c++, java, php, javascript, c\#, \dots                                                                  \\
		Vulnerability/Attack        & sql injection, cross-site scripting, side channel attacks, path traversal, \dots         \\
		Threat Model                & \textless{}any capabilities and restrictions\textgreater{}                                              \\
		Security Feature            & access control, cryptography, input validation                                                          \\
		Code Element                & dataflow, control flow, interprocedural, intraprocedural, static elements                              \\
		Intermediate Representation & control flow graph, dataflow graph, abstract syntax tree, call graph, \dots            \\
		Technique                   & dataflow analysis, control flow analysis, program slicing, pattern matching, \dots   \\
		Check Specification         & hardcoded, pattern-based, query-based, learned                                                          \\
		External Tool               & compiler and analysis frameworks, analyzer, solver and engines, \dots     \\
		Proposes Fix                & yes, no                                                                                                 \\
		Evaluation Metric           & feasibility, performance, precision, recall, f-score, \dots                                               \\
		Evaluation Method           & custom benchmark, case study, systematically created benchmark, \dots \\
		Evaluation Object           & programs, examples, components                                                                          \\
		Evaluation Sample Size      & \textless{}number of evaluation objects\textgreater{}                                                   \\
		Formal Proofs               & soundness, correctness, algorithm properties, complexity, sufficiency                                   \\
		Exploitability              & yes, no                                                                                                 \\
		Confirmed Vulnerability     & yes, no                                                                                                 \\
		Limitations                 & approximations, out of scope, missing features, performance, multi-threading, \dots   				   \\
		\bottomrule
	\end{tabular}
	\vspace{-.2cm}
\end{table}

\looseness=-1
\textbf{Step 4:}
Using open coding, we drew an initial set of categories corresponding to our research questions, and refined the codes based on the analyzed papers.
\Cref{tab:coding_categories} in \cref{sec:domains} below shows the final set of categories and examples of codes.
We divided the papers into 10 batches of 40 papers and one batch of 22 papers.
The first author extracted information from each batch, discussing the extractions with the second author to ensure consistency.
After coding all papers, the first and second author discussed the coding, and merged similar codes into categories, to answer our research questions.
We give an overview over all papers, and show the results of our coding in \Cref{tab:analyzers}. 
We provide a full replication package including filtering decisions for all steps, and the full coding of papers\,\cite{replication}.

%% file: 06_Limitations.tex
\subsection{Threats to Validity}

We discuss potential threats to the validity of our systematic literature review. 

\parhead{Internal Validity.}
The process of title, abstract, and full-paper filtering, as well as open coding, may introduce author bias.
While we independently reviewed initial batches of papers and reported inter-coder agreement (e.g., Cohen's Kappa of up to 0.96 in the abstract filtering phase), some degree of bias may remain, especially for ambiguous cases.
We mitigated this by discussing disagreements, and refining exclusion criteria and our code system.

Many papers did not provide full details on their static analysis techniques.
This may have led to incomplete or imprecise coding of certain categories.
We mitigated this by excluding papers that lacked sufficient information, and by discussing ambiguities collaboratively.

The full-paper review and detailed coding were primarily conducted by the first author.
The first and second authors discussed the results after coding a batch of 40 papers to ensure consistency.

\parhead{External Validity.}
The scope of our search may limit the generalizability of our findings.

Our search strategy was limited to the Scopus digital library, papers written in English, and classified under computer science.
Relevant papers outside this scope may have been missed.
We mitigated this risk by applying a broad search term.

We only considered peer-reviewed journal and conference papers from the computer science domain.
Gray literature, technical reports, or preprints may contain additional relevant static analysis techniques.
Still, we were able to include a large number of papers by using Scopus.

%% file: 04_Results.tex
\section{Static Security Analyzers and their Application Context (RQ1)}
\label{sec:domains}

\Cref{tab:analyzers} gives an overview of all 246 analyzers we identified. The table also provides some of their characteristics, such as the year in which they have been presented, the programming languages they can analyze, and other information extracted for other research questions.

\input{tables/analyzers-longtable.tex}

To determine the application context of all analyzers, we investigated targeted programming languages, application domains, and further objectives. The identified characteristics are as follows.

\subsection{Programming Languages}

\begin{figure}[b]
\vspace{-.2cm}
\begin{minipage}[t]{.44\textwidth}
    \begin{tikzpicture}
        \scriptsize
        \begin{axis}[
            xbar,
            xmin=0,
            xmax=100,
            xlabel=,
            ylabel=,
            axis lines=left,
            enlarge y limits=0.05,
            xmin=0,
            xmajorgrids=true,
            ymajorgrids=false,
            symbolic y coords={
                c/c++,
                java,
                php,
                javascript,
                c\#,
                python,
                rust,
                jvm-compatible,
                lua,
                other,
                unspecified
            },
            ytick=data,
            bar width=0.25cm,
            width=.9\textwidth,
            height=4.75cm,
            y dir=reverse,
            nodes near coords,
            every node near coord/.style={font=\tiny, anchor=east},
            x axis line style={-},
            y axis line style={-},
            point meta=explicit symbolic,
        ]
        \addplot [fill={rgb:red,46; green,92; blue,110}, draw=none,
            every node near coord/.style={anchor=west, font=\tiny ,text=black,opacity=1}] coordinates {
            (99,c/c++)[99]
            (86,java)[86]
            (28,php)[28]
            (15,javascript)[15]
            (7,c\#)[7]
            (6,python)[6]
            (4,rust)[4]
            (4,jvm-compatible)[4]
            (2,lua)[2]
            (8,other)[8]
            (2,unspecified)[2]
        };
        \end{axis}
    \end{tikzpicture} 
\end{minipage}
\begin{minipage}[t]{.55\textwidth}
	\begin{tikzpicture}
		\scriptsize
		\begin{axis}[
			xbar,
			xmin=0,
			xmax=140,
			xlabel=,
			ylabel=,
			axis lines=left,
			enlarge y limits=0.07,
			xmin=0,
			xtick distance=20,
			xmajorgrids=true,
			ymajorgrids=false,
			symbolic y coords={
				domain-independent,
				web applications,
				android,
				operating systems,
				iot,
				cryptography,
				embedded systems,
				distributed systems
			},
			ytick=data,
			bar width=0.25cm,
			width=.85\textwidth,
			height=4cm,
			y dir=reverse,
			nodes near coords,
			every node near coord/.style={font=\tiny, anchor=east},
			x axis line style={-},
			y axis line style={-},
			point meta=explicit symbolic,
			]
			\addplot [fill={rgb:red,46; green,92; blue,110}, draw=none,
			every node near coord/.style={anchor=west, font=\tiny ,text=black,opacity=1}] coordinates {
				(122,domain-independent)[122]
				(49,web applications)[49]
				(36,android)[36]
				(15,operating systems)[15]
				(11,iot)[11]
				(5,cryptography)[5]
				(5,embedded systems)[5]
				(3,distributed systems)[3]
			};
		\end{axis}
	\end{tikzpicture}
\end{minipage}
\vspace{-.8cm}
	\caption{Programming languages and application domains targeted by static security analyzers}
	\label{fig:domains_counts}
	\label{fig:language_counts}
\end{figure}
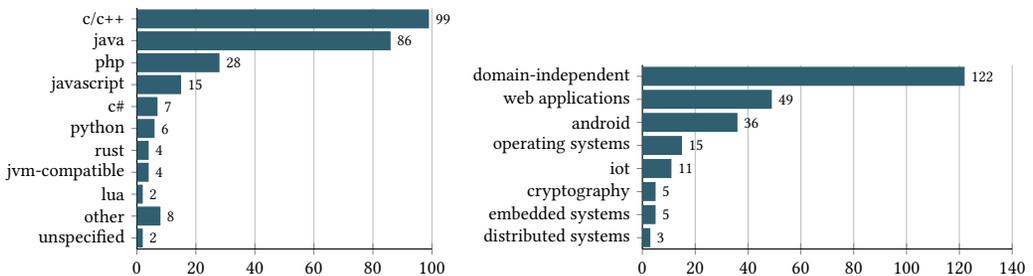

\Cref{fig:language_counts} shows how many security analyzers are capable of analyzing code written in each programming language.
We identified 17 different programming languages.
For 2 analyzers, we could not identify the analyzed programming language, and we therefore categorized them as "unspecified".

The most frequently analyzed programming langauges are C/C++ (99 analyzers) and Java (86 analyzers), which are both general-purpose programming languages widely used in various application domains.
However, PHP (27 analyzers) and JavaScript (15 analyzers) are also frequently analyzed.
Less frequently analyzed programming languages include C\# (7 analyzers), Python (6 analyzers), Rust (4 analyzers), jvm-compatible languages such as Kotlin and Groovy (4 analyzers), and Lua (2 analyzers).
Finally, we identified 8 programming languages that are only analyzed by a single static security analyzer: While, Vb.net, Ur, robotic DSLs, Ansible, Fortran, Erlang, and Go.

\subsection{Application Domains}

We identified 7 different application domains for static security analyzers, which we show in \cref{fig:domains_counts}. 

\emph{Domain-independent} analyzers (116 analyzers) do not specify a domain, and include mostly analyzers for general-purpose programming languages such as C or Java, and detect mainly memory vulnerabilities or improper validation (\cref{sec:vuln}).

Web application analyzers (49 papers) mostly target PHP or JavaScript, but also Java systems.
They typically scan code for injection, cross-site scripting, and improper authentication vulnerabilities.

Analyzers for Android applications (36 analyzers) mostly analyze Java code for vulnerabilities such as improper permission handling, data leakage, and insecure communication.

More sparse domains include operating systems (15 analyzers), which often involve an analysis on the Linux Kernel, and IoT devices (11 analyzers), typically analyzing their firmware.

Embedded systems (5 analyzers), cryptography (5 analyzers), and distributed systems (3 analyzers) are the least targeted domains.
In summary, while most static security analyzers are domain-independent, web applications, Android, operating systems, and IoT devices are frequently targeted.

\subsection{Further Objectives}
\begin{figure}[t]
    \centering
    \begin{tikzpicture}
        \scriptsize
        \begin{axis}[
            xbar,
            xmin=0,
            xmax=11,
            xlabel=,
            ylabel=,
            axis lines=left,
            enlarge y limits=0.12,
            xmin=0,
            xmajorgrids=true,
            ymajorgrids=false,
            symbolic y coords={
                program repair,
                extendability,
                usability,
                process improvements,
                malware detection,
                testing
            },
            ytick=data,
            bar width=0.23cm,
            width=.9\textwidth,
            height=3.4cm,
            y dir=reverse,
            nodes near coords,
            every node near coord/.style={font=\tiny, anchor=east},
            x axis line style={-},
            y axis line style={-},
            point meta=explicit symbolic,
        ]
        \addplot [fill={rgb:red,46; green,92; blue,110}, draw=none,
            every node near coord/.style={anchor=west, font=\tiny ,text=black,opacity=1}] coordinates {
            (10,program repair)[10]
            (6,extendability)[6]
            (5,usability)[5]
            (4,process improvements)[4]
            (3,malware detection)[3]
            (2,testing)[2]
        };
        \end{axis}
    \end{tikzpicture}
    \vspace{-.3cm}
    \caption{Further objectives of static security analyzers}
    \label{fig:objectives_counts}
    \vspace{-.3cm}
\end{figure}
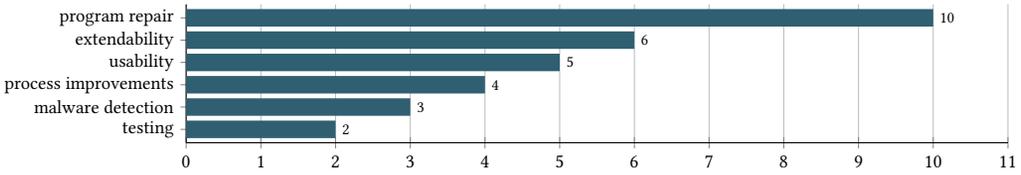

Although the primary objective of static security analyzers is to detect vulnerabilities in source code, some papers also mention further objectives, shown in \cref{fig:objectives_counts}.
\emph{Program repair} (10 analyzers) aims to automatically fix detected vulnerabilities.
In addition, we found that only 5.7\,\% of the static security analyzers propose fixes for detected vulnerabilities, while 94.3\,\% do not.
Six static security analyzers are also designed to be \emph{extendable}, allowing users to easily add new checks or rules to the analyzer.
Other static security analyzers focus on software \emph{process improvements} (4 analyzers), such as being able to integrate static security analysis into code implementation workflows, or \emph{usability} (5 analyzers).
Although we excluded analyzers that primarily focus on \emph{malware detection} during our selection process, 3 papers mention this as a further objective of their static security analyzer.
Finally, \emph{testing} (2 analyzers) is also listed by some papers as a further objective, aiming to generate test cases or malicious inputs that can be used to test for vulnerabilities.

\quotebox{Static analyzer domains (RQ1)}{Static security analyzers are typically multi-purpose, predominantly targeting C/C++ and Java programs not specific to a domain. However, domains such as web applications, Android, operating systems, or IoT are frequently targeted. Some analyzers additionally aim for program repair, extendability, process improvement, usability, malware detection, or testing.}{5cm}

\section{Weaknesses Detected by Static Security Analyzers (RQ2)}
\label{sec:vuln}
To understand identifiable vulnerabilities, we examined the weaknesses supported by the analyzers.

\subsection{Comprehensive Categorization of Weaknesses}

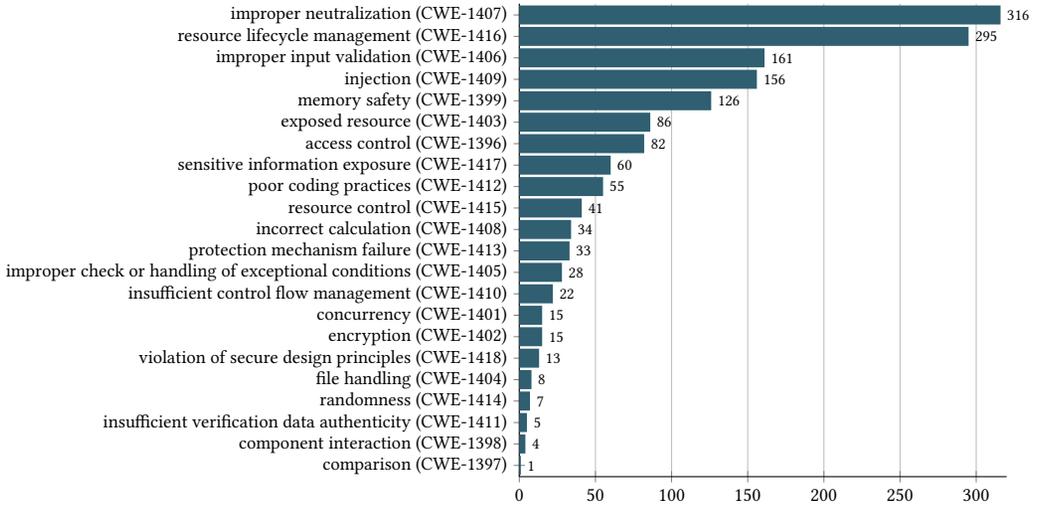
\begin{figure}[t]
    \centering
    \begin{tikzpicture}
    \scriptsize
    \begin{axis}[
        xbar,
        xmin=0,
        xmax=320,
        axis lines=left,
        enlarge y limits=0.025,
        xmajorgrids=true,
        ymajorgrids=false,
        symbolic y coords={
            c1407,c1416,c1406,c1409,c1399,c1403,c1396,c1417,c1412,c1415,
            c1408,c1413,c1405,c1410,c1401,c1402,c1418,c1404,c1414,c1411,
            c1398,c1397
        },
        ytick=data,
        yticklabels={
            improper neutralization (CWE-1407),
            resource lifecycle management (CWE-1416),
            improper input validation (CWE-1406),
            injection (CWE-1409),
            memory safety (CWE-1399),
            exposed resource (CWE-1403),
            access control (CWE-1396),
            sensitive information exposure (CWE-1417),
            poor coding practices (CWE-1412),
            resource control (CWE-1415),
            incorrect calculation (CWE-1408),
            protection mechanism failure (CWE-1413),
            improper check or handling of exceptional conditions (CWE-1405),
            insufficient control flow management (CWE-1410),
            concurrency (CWE-1401),
            encryption (CWE-1402),
            violation of secure design principles (CWE-1418),
            file handling (CWE-1404),
            randomness (CWE-1414),
            insufficient verification data authenticity (CWE-1411),
            component interaction (CWE-1398),
            comparison (CWE-1397)
        },
        bar width=0.25cm,
        width=.58\textwidth,
        height=7.8cm,
        y dir=reverse,
        nodes near coords,
        every node near coord/.style={font=\tiny, anchor=east},
        x axis line style={-},
        y axis line style={-},
        point meta=explicit symbolic,
    ]

    \addplot[fill={rgb:red,46; green,92; blue,110}, draw=none,
        every node near coord/.style={anchor=west, font=\tiny, text=black}] coordinates {
        (316,c1407)[316]
        (295,c1416)[295]
        (161,c1406)[161]
        (156,c1409)[156]
        (126,c1399)[126]
        (86,c1403)[86]
        (82,c1396)[82]
        (60,c1417)[60]
        (55,c1412)[55]
        (41,c1415)[41]
        (34,c1408)[34]
        (33,c1413)[33]
        (28,c1405)[28]
        (22,c1410)[22]
        (15,c1401)[15]
        (15,c1402)[15]
        (13,c1418)[13]
        (8,c1404)[8]
        (7,c1414)[7]
        (5,c1411)[5]
        (4,c1398)[4]
        (1,c1397)[1]
    };

    \end{axis}
    \end{tikzpicture}
	\vspace{-.3cm}
    \caption{Weaknesses scanned by static security analyzers organized by top-level CWE comprehensive categories}
    \label{fig:vulnerability_counts}
    \vspace{-.3cm}
\end{figure}

We used weaknesses from the CWE as an abstraction to categorize detection capabilities, since many analyzer directly map to them.
We used three sources to identify CWEs supported by each analyzer: (1) explicitly mentioned CWEs, (2) mentioned CVEs, and (3) mentioned vulnerabilities and attacks in the papers, which we manually mapped to CWEs.
Each CWE is related to other CWEs through parent\/child-relationships, which we used to lift detected weaknesses to a higher abstraction level for better comprehension.
As the highest abstraction level, we used the comprehensive categorization for software assurance trends,\footnote{https://cwe.mitre.org/data/definitions/1400.html} which assigns each CWE to one category of weaknesses.
Still, due to their relationship with each other, some weaknesses may be children of multiple weaknesses which fall into more than one category.
For example, the extracted CWEs from category CWE-1406 for improper input validation are also fully contained in CWE-1407 for improper neutralization, which results in a strong overlap between some of the CWE categories.
\Cref{fig:vulnerability_counts} lists the weaknesses detected by static security analyzers, and \Cref{tab:analyzers} gives a full overview for each analyzer.

Static security analyzers related to \emph{improper neutralization} (316 weaknesses) check that input is well-formed and and validated by transforming the input, or rejecting malicious input.

Most static security analyzers for \emph{resource lifecycle management} (295 weaknesses) check for improper control of resources through its lifetime (CWE-664), which include the use of objects before it is created, or after it is destroyed.

\emph{Improper input validation} (161 weaknesses) is a form of improper neutralization, in which a software system does not correctly check potentially dangerous inputs to ensure safe execution.
The system could then be vulnerable to \emph{injection} attacks (156 weaknesses), in which an attacker provides a malicious command or data structure which is not properly sanitized from special elements that could be interpreted by the system.

Furthermore, static security analyzers are often concerned with \emph{memory safety} (126 weaknesses).
Buffer overflows (CWE-119) are the most targeted weakness in this category, and describe read or write operations outside the intended boundaries of a memory buffer.
Use after free and double free are common examples, as they access a memory address after it has been freed.

An \emph{exposed resource} (86 weaknesses) is a kind of resource lifecycle management issue that allows unauthorized actors to access sensitive information (CWE-200).
In this case, the information flows into e.g., variables or messages that are accessible to such actors.
CWE-200 is also included in the category \emph{sensitive information exposure} (60 weaknesses).

Weaknesses related to \emph{access control} (82 weaknesses) are often related to improper authentication (CWE-287), which is used to prove the identity of an actor, or improper authorization (CWE-285), which restricts access to a resource to a given actor.
Such weaknesses include the use of hard-coded credentials, or missing access control checks.

\emph{Poor coding practices} (55 weaknesses) are insecure coding patterns like using potentially dangerous functions (CWE-676), improper following of specifications (CWE-573), or dead code (CWE-561).

Fewer weaknesses fall into the categories of \emph{resource control} (41 weaknesses), \emph{incorrect calculation} (34 weaknesses), \emph{protection mechanism failure} (33 weaknesses), \emph{improper check or handling of exceptional conditions} (28 weaknesses), \emph{insufficient control flow management} (22 weaknesses), \emph{encryption} (15 weaknesses), or \emph{concurrency} (15 weaknesses).

The smallest categories comprise weaknesses related to \emph{violation of secure design principles} (13 weaknesses), \emph{file handling} (8 weaknesses), \emph{randomness} (7 weaknesses), \emph{insufficient verification of data authenticity} (5 weaknesses), \emph{component interaction} (4 weaknesses), and \emph{comparison} (1 weakness).

While static security analyzers cover some categories of weaknesses extensively, other categories are only sparsely covered.
One cause for the distribution of weaknesses is that some categories cover a wider range of weaknesses than others.
For example, while improper neutralization covers 68 different CWEs, incorrect calculation only covers 12 CWEs.

\subsection{Low-level Weaknesses}

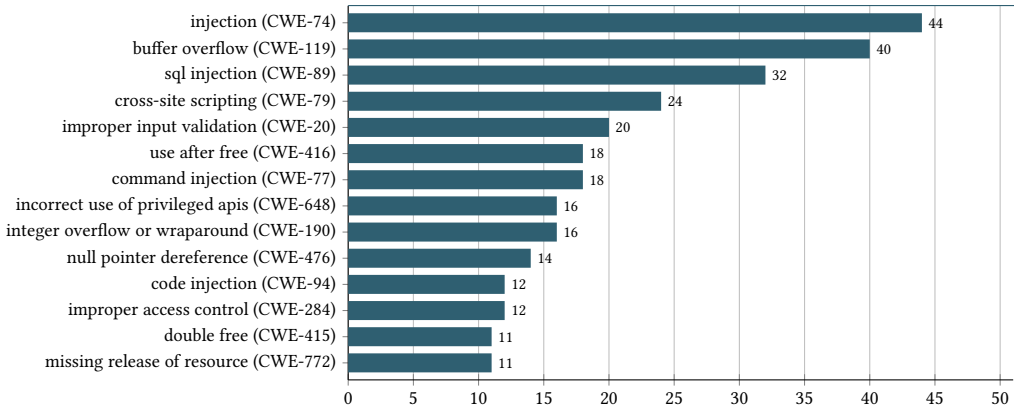
\begin{figure}[t]
    \centering
    \begin{tikzpicture}
    \scriptsize
    \begin{axis}[
        xbar,
        xmin=0,
        xmax=51,
        axis lines=left,
        enlarge y limits=0.05,
        xmajorgrids=true,
        ymajorgrids=false,
        symbolic y coords={
            C200,
            C74,
            C119,
            C89,
            C79,
            C20,
            C416,
            C77,
            C648,
            C190,
            C476,
            C94,
            C284,
            C415,
            C772
        },
        ytick=data,
        yticklabels={
            exposure of sensitive information (CWE-200),
            injection (CWE-74),
            buffer overflow (CWE-119),
            sql injection (CWE-89),
            cross-site scripting (CWE-79),
            improper input validation (CWE-20),
            use after free (CWE-416),
            command injection (CWE-77),
            incorrect use of privileged apis (CWE-648),
            integer overflow or wraparound (CWE-190),
            null pointer dereference (CWE-476),
            code injection (CWE-94),
            improper access control (CWE-284),
            double free (CWE-415),
            missing release of resource (CWE-772)
        },
        bar width=0.25cm,
        width=.75\textwidth,
        height=6.5cm,
        y dir=reverse,
        nodes near coords,
        every node near coord/.style={font=\tiny, anchor=east},
        x axis line style={-},
        y axis line style={-},
        point meta=explicit symbolic,
    ]

    \addplot[fill={rgb:red,46; green,92; blue,110}, draw=none,
        every node near coord/.style={anchor=west, font=\tiny, text=black}] coordinates {
        (52,C200)[52]
        (44,C74)[44]
        (40,C119)[40]
        (32,C89)[32]
        (24,C79)[24]
        (20,C20)[20]
        (18,C416)[18]
        (18,C77)[18]
        (16,C648)[16]
        (16,C190)[16]
        (14,C476)[14]
        (12,C94)[12]
        (12,C284)[12]
        (11,C415)[11]
        (11,C772)[11]
    };

    \end{axis}
    \end{tikzpicture}
    \vspace{-.35cm}
    \caption{Weaknesses scanned by static security analyzers}
    \label{fig:cwe_counts_concrete}
    \vspace{-.15cm}
\end{figure}

To gain a deeper understanding of weaknesses detected by static security analyzers, we examined the most frequent CWEs.
\Cref{fig:cwe_counts_concrete} shows the 15 most frequently detected CWEs by static security analyzers.
We found that on average, static security analyzers detect 2.75 different CWEs.

The most frequently detected weakness is the \emph{exposure of sensitive information to an unauthorized actor} (CWE-200) (52 analyzers), in which an actor gains access to information without authorization.

\emph{Injection} (CWE-74) (44 analyzers) allows an attacker to supply untrusted input to a program, which is then executed as part of a command or query.
\emph{SQL injection} (CWE-89) (32 analyzers), \emph{command injection} (CWE-77) (18 analyzers), and \emph{code injection} (CWE-94) (12 analyzers) are specific types of injection that target SQL databases, operating system commands, and application source code, respectively.
Injection attacks are often caused by \emph{improper input validation} (CWE-20) (20 analyzers), in which a software system does not correctly check potentially dangerous inputs.
This can also lead to \emph{cross-site scripting} (CWE-79) (24 analyzers), in which an attacker injects malicious inputs, e.g., into web applications that are then executed in the context of a user's browser.

Memory safety issues are also frequently detected by static security analyzers.
\emph{Improper restriction of operations within the bounds of a memory buffer} (CWE-119) (40 analyzers) is a weakness that allows an attacker to read or write outside the intended boundaries of a memory buffer.
\emph{Use after free} (CWE-416) (18 analyzers) and \emph{double free} (CWE-415) (11 analyzers) are types of memory safety weaknesses that occur when a software systems uses or frees a memory address after it has been freed.
Similarly, \emph{missing release of resource after effective lifetime} (CWE-772) (11 analyzers) allows an attacker to exhaust system resources by not releasing when they are not needed anymore.
An \emph{integer overflow or wraparound} (CWE-190) (16 analyzers) is caused by a calculation that exceeds the maximum size of an integer data type.
\emph{Null pointer dereference} (CWE-476) (14 analyzers) occurs when a program attempts to use a pointer that has a null value.

Finally, \emph{incorrect use of privileged APIs} (CWE-648) (16 analyzers) and \emph{improper access control} (CWE-284) (12 analyzers) are weaknesses related to access control issues, which allow an attacker to gain unauthorized access to resources or perform actions that should be restricted.

In summary, static security analyzers address weaknesses related to data exposure, injection attacks, memory safety issues, and access control flaws.

\subsection{Security Features}

\begin{figure}[t]
    \centering
    \begin{tikzpicture}
        \scriptsize
        \begin{axis}[
            xbar,
            xmin=0,
            xmax=30,
            xlabel=,
            ylabel=,
            axis lines=left,
            enlarge y limits=0.25,
            xmin=0,
            xmajorgrids=true,
            ymajorgrids=false,
            symbolic y coords={
                access control,
                cryptography,
                input validation
            },
            ytick=data,
            bar width=0.25cm,
            width=.95\textwidth,
            height=2.5cm,
            y dir=reverse,
            nodes near coords,
            every node near coord/.style={font=\tiny, anchor=east},
            x axis line style={-},
            y axis line style={-},
            point meta=explicit symbolic,
        ]
        \addplot [fill={rgb:red,46; green,92; blue,110}, draw=none,
            every node near coord/.style={anchor=west, font=\tiny ,text=black,opacity=1}] coordinates {
            (26,access control)[26]
            (13,cryptography)[13]
            (7,input validation)[7]
        };
        \end{axis}
    \end{tikzpicture}
    \vspace{-.3cm}
    \caption{Security features related to static security analyzers}
    \label{fig:defenses_counts}
    \vspace{-.1cm}
\end{figure}
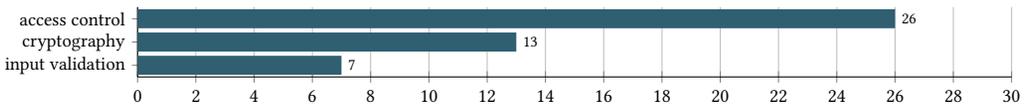

While static security analyzers primarily target general weaknesses, some weaknesses such as inadequate encryption strength (CWE-326) or improper access control (CWE-284) are directly related to security features.
Security features are functionalities to protect a software system against attacks, and are mentioned in 46 papers, shown in \cref{fig:defenses_counts}.
\emph{Access control} (24 papers) is the most frequently mentioned security feature, and is typically mentioned in the context of analyzing improper authorization or authentication vulnerabilities.
Such analyzers often check whether access control checks are properly implemented before accessing sensitive resources, or whether access control policies are correctly enforced.
Besides access control, \emph{cryptography} (13 papers) is mentioned in the context of analyzing improper use of cryptographic libraries, use of securely random generated numbers, or checking for vulnerabilities in cryptographic algorithms.
Finally, \emph{input validation} (7 papers) is mentioned in the context of analyzing whether user inputs are properly sanitized and validated to prevent injection attacks or other vulnerabilities.
Although, security features are rarely mentioned in the context of static security analyzers, many analyzers could implicitly analyze code related to such features when checking for vulnerabilities, e.g., analyzing cryptography when checking for weak encryption or insecure api usage.

\quotebox{Vulnerabilities (RQ2)}{Static security analyzers primarily target vulnerabilities related to improper neutralization, resource lifecycle management, and improper input validation, which include injection attacks and memory safety issues such as buffer overflows and use-after-free. They also frequently address exposed resources, access control flaws, and sensitive information exposure. Less frequently, analyzers cover weaknesses related to security features like access control, or cryptography.}{5cm}

\section{Analysis Techniques Employed by Static Security Analyzers (RQ3)}
We examined the elements analyzed by static security tools, along with their intermediate representations, analysis techniques, check specifications, and any external tools they rely on.

\subsection{Analyzed Flows and Code Elements}

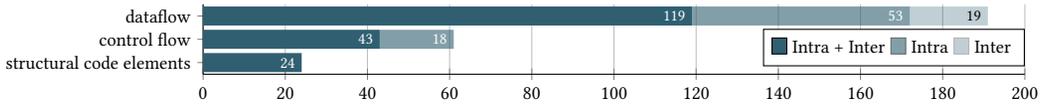
\begin{figure}[t]
    \centering
    \begin{tikzpicture}
        \scriptsize
        \begin{axis}[
            xbar stacked,
            xmin=0,
            xmax=200,
            xlabel=,
            ylabel=,
            axis lines=left,
            enlarge y limits=0.25,
            xmajorgrids=true,
            ytick=data,
            bar width=0.25cm,
            width=.9\textwidth,
            height=2.5cm,
            y dir=reverse,
            symbolic y coords={
                dataflow,
                control flow,
                structural code elements
            },
            nodes near coords,
            every node near coord/.style={font=\tiny, anchor=east},
            x axis line style={-},
            y axis line style={-},
            point meta=explicit symbolic,
            legend style={
                at={(1,0.65)},
                anchor=north east,
                draw=black,
                fill=white,
                font=\scriptsize,
                legend columns=3,
                legend cell align=left
            },
        ]

        \addplot [fill={rgb:red,46; green,92; blue,110}, draw=none,
            every node near coord/.style={anchor=east, font=\tiny ,text=white,opacity=1}] coordinates {
            (119,dataflow)[119]
            (43,control flow)[43]
            (24,structural code elements)[24]
        };
        \addlegendentry{Intra + Inter}
        \addplot [fill={rgb:red,46; green,92; blue,110}, opacity=0.6, draw=none,
            every node near coord/.style={anchor=east, font=\tiny ,text=white,opacity=1}] coordinates {
            (53,dataflow)[53]
            (18,control flow)[18]
            (0,structural code elements)[]
        };
        \addlegendentry{Intra}
        \addplot [fill={rgb:red,46; green,92; blue,110}, opacity=0.3, draw=none,
            every node near coord/.style={anchor=east, font=\tiny ,text=black,opacity=1}] coordinates {
            (19,dataflow)[19]
            (0,control flow)[]
            (0,structural code elements)[]
        };
        \addlegendentry{Inter}

        \end{axis}
    \end{tikzpicture}
    \vspace{-.8cm}
    \caption{Code elements analyzed by static security analyzers}
    \label{fig:elements_counts}
    \vspace{-.1cm}
\end{figure}

Static security analyzers examine dataflow, control flow, or structural code elements such as hardcoded strings.
Flows can be intraprocedural (within a single procedure or component) or interprocedural (across multiple procedures or components).
\Cref{fig:elements_counts} shows the number of analyzers for each flow type and structural element, and \Cref{tab:analyzers} gives a full overview for each analyzer.

The majority (191 analyzers) analyzes \emph{dataflow}, tracking the flow of data through variables, functions, or objects to identify vulnerabilities such as injection attacks, or data leakage. 
Analyzers mostly consider both \emph{intra - and interprocedural} (119 analyzers) dataflows.
However, 53 analyzers only consider \emph{intraprocedural} dataflows, in which calls to other procedures are either ignored or approximated.
Only few analyzers exclusively consider \emph{interprocedural} (19 analyzers) dataflows, which often analyze call graphs to track dataflow across multiple components. 

\emph{Control flow} (60 analyzers) captures the execution order of instructions or statements. 
Similar to dataflow, most analyzers consider both \emph{intra - and interprocedural} (43 analyzers) control flow.
Still, 18 analyzers only consider \emph{intraprocedural} control flow, no analyzer exclusively considers \emph{interprocedural} control flow.
In addition to control flow, 22 analyzers analyze dataflow.

Finally, 24 analyzers analyze \emph{structural code elements} such as specific call statements, variable declarations, or code patterns to identify e.g., dangerous functions, or hard-coded credentials. 

\subsection{Intermediate Representations}

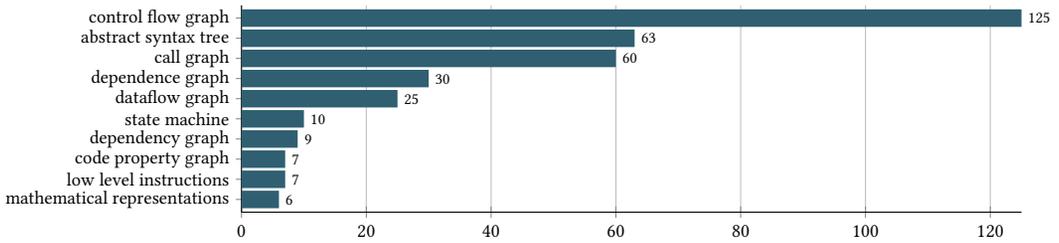
\begin{figure}[t]
    \centering
    \begin{tikzpicture}
        \scriptsize
        \begin{axis}[
            xbar,
            xmin=0,
            xmax=125,
            xlabel=,
            ylabel=,
            axis lines=left,
            enlarge y limits=0.07,
            xmin=0,
            xmajorgrids=true,
            ymajorgrids=false,
            symbolic y coords={
                control flow graph,
                abstract syntax tree,
                call graph,
                dependence graph,
                dataflow graph,
                state machine,
                dependency graph,
                code property graph,
                low level instructions,
                mathematical representations
            },
            ytick=data,
            bar width=0.23cm,
            width=.86\textwidth,
            height=4.3cm,
            y dir=reverse,
            nodes near coords,
            every node near coord/.style={font=\tiny, anchor=east},
            x axis line style={-},
            y axis line style={-},
            point meta=explicit symbolic,
        ]
        \addplot [fill={rgb:red,46; green,92; blue,110}, draw=none,
            every node near coord/.style={anchor=west, font=\tiny ,text=black,opacity=1}] coordinates {
            (125,control flow graph)[125]
            (63,abstract syntax tree)[63]
            (60,call graph)[60]
            (30,dependence graph)[30]
            (25,dataflow graph)[25]
            (10,state machine)[10]
            (9,dependency graph)[9]
            (7,code property graph)[7]
            (7,low level instructions)[7]
            (6,mathematical representations)[6]
        };
        \end{axis}
    \end{tikzpicture}
    \vspace{-.8cm}
    \caption{Intermediate representations used by static security analyzers}
    \label{fig:transformation_counts}
    \vspace{-.4cm}
\end{figure}

Intermediate representations serve as an abstraction of source code and are reported 347 times in static security analyzer papers, which we grouped into 10 categories as shown in \cref{fig:transformation_counts}.

\emph{Control flow graphs} (CFG)\,\cite{allen1970} (125 analyzers) represent the flow of control within a program, with nodes representing basic blocks of code, and edges representing control flow between these blocks.
CFGs are used for analyses that require an understanding of the program's execution order. 

\emph{Abstract syntax tree} (AST)\,\cite{alfred2007} (60 analyzers) represent the hierarchical structure of source code, with nodes containing language constructs (e.g., expressions, statements, declarations), and edges describing their relationship.
While some analyses operate on ASTs, they are often used to derive other representations such as CFGs and may therefore not be explicitly mentioned in all papers.

\emph{Call graphs}\,\cite{alfred2007} (60 analyzers) describe inter-procedural control flow between functions or methods, where nodes encompass functions and edges represent function calls. 

\emph{Dependence graphs}\,\cite{Ferrante1987} (30 analyzers) show data and control dependencies between statements.

\emph{dataflow graphs} (DFGs)\,\cite{Orailoglu1986} (25 analyzers) model the flow of data within a program, with nodes representing variables or data structures and edges representing data dependencies. 

\emph{State machines}\,\cite{harel1987} (10 analyzers) model the behavior of a system as a set of states and transitions between those states based on events, actions or conditions.

\emph{Dependency graphs} (9 analyzers) represent imported modules and their interactions in a program.

\emph{Code property graphs} (CPG)\,\cite{Yamaguchi2014Modeling} (7 analyzers) merge CFG, AST, and DFG into one representation.

Less frequently used intermediate representations include \emph{low level instructions} (7 analyzers), such as three address code, and \emph{mathematical representations} (6 analyzers), such as boolean formulas.

\subsection{Analysis Techniques}

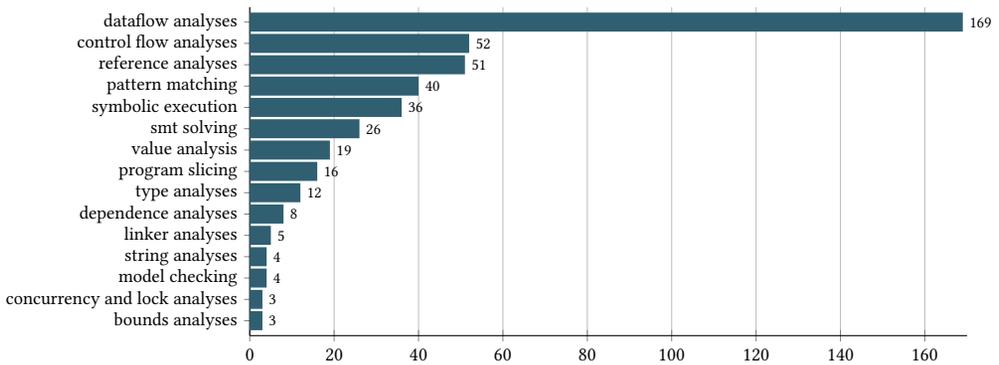
\begin{figure}[t]
    \centering
    \begin{tikzpicture}
        \scriptsize
        \begin{axis}[
            xbar,
            xmin=0,
            xmax=170,
            xlabel=,
            ylabel=,
            axis lines=left,
            enlarge y limits=0.05,
            xmin=0,
            xmajorgrids=true,
            ymajorgrids=false,
            symbolic y coords={
                dataflow analyses,
                control flow analyses,
                reference analyses,
                pattern matching,
                symbolic execution,
                smt solving,
                value analysis,
                program slicing,
                type analyses,
                dependence analyses,
                linker analyses,
                string analyses,
                model checking,
                concurrency and lock analyses,
                bounds analyses
            },
            ytick=data,
            bar width=0.25cm,
            width=.8\textwidth,
            height=5.9cm,
            y dir=reverse,
            nodes near coords,
            every node near coord/.style={font=\tiny, anchor=east},
            x axis line style={-},
            y axis line style={-},
            point meta=explicit symbolic,
        ]
        \addplot [fill={rgb:red,46; green,92; blue,110}, draw=none,
            every node near coord/.style={anchor=west, font=\tiny ,text=black,opacity=1}] coordinates {
            (169,dataflow analyses)[169]
            (52,control flow analyses)[52]
            (51,reference analyses)[51]
            (40,pattern matching)[40]
            (36,symbolic execution)[36]
            (26,smt solving)[26]
            (19,value analysis)[19]
            (16,program slicing)[16]
            (12,type analyses)[12]
            (8,dependence analyses)[8]
            (5,linker analyses)[5]
            (4,string analyses)[4]
            (4,model checking)[4]
            (3,concurrency and lock analyses)[3]
            (3,bounds analyses)[3]
        };
        \end{axis}
    \end{tikzpicture}
    \vspace{-.4cm}
    \caption{Analysis techniques used by static security analyzers}
    \label{fig:techniques_counts}
    \vspace{-.2cm}
\end{figure}

We identified 16 categories of analysis techniques as shown in \cref{fig:techniques_counts}.
\Cref{tab:analyzers} lists for each static security analyzer all techniques used by that analyzer.

\emph{Dataflow analyses} (169 analyzers) mostly encompass taint analysis to track the flow of data to identify injection vulnerabilities, or dataflow violations.
They check whether data from a source (e.g., user input) can reach a sink (e.g., a database query) without sanitization or validation.

\emph{Control flow analyses} (52 analyzers) examine the execution order of instructions or statements in a program to identify vulnerabilities such as infinite loops, or improper exception handling. 

\emph{Reference analyses} (51 analyzers) track references to variables, objects, or memory locations to identify vulnerabilities such as use-after-free, or null pointer dereferences. 
They often involve pointer analysis to determine the possible values that pointers can reference. 

\emph{Pattern matching} (40 analyzers) identifies known vulnerability patterns in the source code.
These patterns can be based on common coding mistakes, insecure coding practices, or known vulnerability signatures.
Pattern matching is often performed using regular expressions graph patterns on abstract syntax trees or graph representations.

\emph{Symbolic execution} (36 analyzers) explores all execution paths of a program by treating inputs as symbolic values rather than concrete values.
It identifies vulnerabilities such as buffer overflows by systematically exploring different input combinations and program states.

Similarly, \emph{SMT solving} (26 analyzers) reasons about the satisfiability of logical formulas representing program properties.
It checks for the existence of inputs that can lead to specific program states or behaviors, such as reaching a vulnerable state or violating a security policy.

\emph{Value analysis} (19 analyzers) tracks the possible values that variables can take at different points in a program.
This technique identifies vulnerabilities such as out-of-bounds accesses, division by zero, or invalid type casts by analyzing whether variables can assume malicious values.

\emph{Program slicing} (16 analyzers) extracts code that affects a specific computation or variable by applying a slicing criterion, such as a variable or program point, to isolate specific portions.

\emph{Type analyses} (12 analyzers) verify that a program adheres to type safety rules, identifying vulnerabilities such as type confusion, invalid casts, or unsafe type conversions.

\emph{Dependence analyses} (8 analyzers) analyzes the dependence between different control or data elements for vulnerability detection, such as whether a variable is influenced by a user input or whether a function call depends on a specific condition.

Less frequently used analysis techniques include \emph{linker analyses} (5 papers) that analyze dependencies between program elements, \emph{model checking} (4 papers) that verify properties of models representing the program, \emph{string analyses} (4 papers) that identify whether strings may assume vulnerable values, \emph{concurrency and lock analyses} (3 papers) that identify issues related to concurrent execution and synchronization, and or \emph{bounds analyses} (3 papers) that check for violations of specified bounds or limits on data structures such as arrays or integers.

\subsection{Check Specifications}

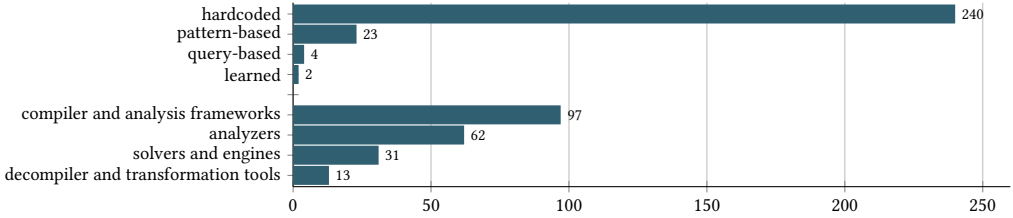
\begin{figure}[t]
    \centering
    \begin{tikzpicture}
        \scriptsize
        \begin{axis}[
            xbar,
            xmin=0,
            xmax=260,
            xlabel=,
            ylabel=,
            axis lines=left,
            enlarge y limits=0.07,
            xmin=0,
            xmajorgrids=true,
            ymajorgrids=false,
            symbolic y coords={
                hardcoded,
                pattern-based,
                query-based,
                learned,
                blank,
                compiler and analysis frameworks,
                analyzers,
                solvers and engines,
                decompiler and transformation tools
            },
            yticklabels={
                hardcoded,
                pattern-based,
                query-based,
                learned,
                ,
                compiler and analysis frameworks,
                analyzers,
                solvers and engines,
                decompiler and transformation tools
            },
            ytick=data,
            bar width=0.25cm,
            width=.8\textwidth,
            height=4cm,
            y dir=reverse,
            nodes near coords,
            every node near coord/.style={font=\tiny, anchor=east},
            x axis line style={-},
            y axis line style={-},
            point meta=explicit symbolic,
        ]
        \addplot [fill={rgb:red,46; green,92; blue,110}, draw=none,
            every node near coord/.style={anchor=west, font=\tiny ,text=black,opacity=1}] coordinates {
            (240,hardcoded)[240]
            (23,pattern-based)[23]
            (4,query-based)[4]
            (2,learned)[2]
            (0,blank)[]
            (97,compiler and analysis frameworks)[97]
            (62,analyzers)[62]
            (31,solvers and engines)[31]
            (13,decompiler and transformation tools)[13]
        };
        \end{axis}
    \end{tikzpicture}
    \vspace{-.3cm}
    \caption{Specifications and external tools used by static security analyzers}
    \label{fig:specification_counts}
    \vspace{-.2cm}
\end{figure}

Static security analyzers use four types of check specifications, shown in \cref{fig:specification_counts}.
Nearly all checks of static security analyzers (240 papers) are \emph{hardcoded} directly in the source code of the analyzer.
While a few static security analyzers use constructs from other specification types, they are either hardcoded in the analyzers' source code, or it is not explicitly mentioned whether they can be modified or extended.
\emph{Pattern-based} (23 papers) specifications define rules or are written in domain-specific languages (DSLs)\,\cite{krausz2024120,dslbook} to describe characteristics of vulnerabilities.
\emph{Query-based} (4 papers) specifications use query languages such as Datalog to express the checks. 
Finally, 2 static security analyzers use specifications \emph{learned} derived from mined CVEs.

\subsection{External Tools}


Static security analyzers often build upon or extend existing tools and frameworks to leverage their capabilities and avoid custom implementations of functionalities.
External tools were mentioned a total of 222 times in 159 papers (1.4 tools on average), which we grouped into 4 categories as shown in \cref{fig:specification_counts}.
A total of 79 papers did not mention any external tools.

\emph{Compiler and analysis frameworks} (97 papers) provide a foundation for building static analysis tools by offering functionalities such as parsing, intermediate representations, and analysis APIs.
The most frequent examples are LLVM (33 times), Soot (33 times), and WALA (11 times).

Static security analyzers often build upon or extend other \emph{analyzers} (62 papers).
We found a wide variety of analyzers, with Julia (4 mentions) and Pixy (3 mentions) being the most frequently mentioned ones, and many analyzers only mentioned once or twice.

\emph{Solvers and engines} (31 papers) are used to solve specific problems or perform specific tasks within the analysis, such as constraint solving or symbolic execution.
Z3 (18 mentions) and the neo4j graph database (5 mentions) are the most frequently mentioned examples in this category.


Finally, \emph{decompiler and transformation tools} (13 papers) are used to transform source code or binaries into a different language or code representation such as bytecode.

\quotebox{Techniques (RQ3)}{Most static security analyzers examine dataflow, often combining intra- and interprocedural analysis to track how information propagates through software. They typically use control flow graphs, abstract syntax trees, and call graphs, to model program structure or behaviour. They employ diverse analysis techniques, with dataflow analysis, reference analysis, and control flow analysis being the most frequent, alongside symbolic execution, SMT solving, and pattern matching. To this end, they leverage external tools such as compiler frameworks, solvers, or other analyzers. Checks are typically hardcoded, and rare use extendable patterns, queries, or learned specifications.}{5cm}

\section{Evaluation Practices of Static Security Analyzers (RQ4)}
To investigate the evaluation practices of static security analyzers, we recorded the evaluation metrics, benchmarks, and sample sizes used in the papers, as well as any provided formal proofs.

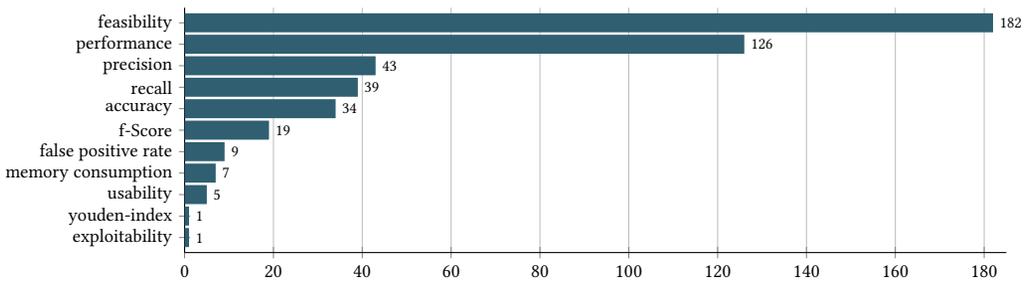
\begin{figure}[t]
    \centering
    \begin{tikzpicture}
        \scriptsize
        \begin{axis}[
            xbar,
            xmin=0,
            xmax=185,
            xlabel=,
            ylabel=,
            axis lines=left,
            enlarge y limits=0.07,
            xmin=0,
            xmajorgrids=true,
            ymajorgrids=false,
            symbolic y coords={
                feasibility,
                performance,
                precision,
                recall,
                accuracy,
                f-Score,
                false positive rate,
                memory consumption,
                usability,
                youden-index,
                exploitability
            },
            ytick=data,
            bar width=0.25cm,
            width=.9\textwidth,
            height=4.8cm,
            y dir=reverse,
            nodes near coords,
            every node near coord/.style={font=\tiny, anchor=east},
            x axis line style={-},
            y axis line style={-},
            point meta=explicit symbolic,
        ]
        \addplot [fill={rgb:red,46; green,92; blue,110}, draw=none,
            every node near coord/.style={anchor=west, font=\tiny ,text=black,opacity=1}] coordinates {
            (182,feasibility)[182]
            (126,performance)[126]
            (43,precision)[43]
            (39,recall)[39]
            (34,accuracy)[34]
            (19,f-Score)[19]
            (9,false positive rate)[9]
            (7,memory consumption)[7]
            (5,usability)[5]
            (1,youden-index)[1]
            (1,exploitability)[1]
        };
        \end{axis}
    \end{tikzpicture}
    \vspace{-.3cm}
    \caption{Metrics used to evaluate static security analyzers}
    \label{fig:metrics_counts}
    \vspace{-.2cm}
\end{figure}

\subsection{Evaluation Metrics}

The primary objective of static security analyzers is detecting vulnerabilities in source code.
To evaluate their effectiveness in doing so, authors of such analyzers use 11 metrics, illustrated in \cref{fig:metrics_counts}.
We grouped similar metrics, which are calculated using the same measurements, such as recall, sensitivity, and true positive rate into single groups.

\emph{Feasibility} (182 papers) shows that the proposed static analysis tool is able to detect vulnerabilities without using any kind of statistical measurement besides a count of detected vulnerabilities.
Typically, the evaluation includes selecting some software systems, often involving custom benchmarks, and using them as an input for the analyzer.
The evaluation then entrails a discussion of the reports of the analyzer, which presents true and false positives along with some examples.

\emph{Performance} (126 papers) measures the runtime of the static security analysis tool.

\emph{Precision} (43 papers) is the relation of the number of true positives to the number of true and false positives, whereas \emph{recall} (39 papers) substitutes the false positives with false negatives.
\emph{Accuracy} (34 papers) weighs the relation of true positives and negatives to all findings.
The \emph{F-Score} (19 papers) represents the harmonic mean of precision and recall.
The \emph{false positive rate} (9 papers) measures the relation between false positives and the number of actual negatives.
These metrics require measuring the number of true and false positives as well as negatives.
However, such numbers are only available when using a labeled benchmark, and unknown for custom benchmark without a ground truth, which could be the cause for the lower number of papers using these metrics.

Sporadically, some papers measure the \emph{Memory consumption} (7 papers) of the analysis, \emph{usability} (5 papers), the \emph{youden-index} (1 paper), and the \emph{exploitability} (1 paper) of detected vulnerabilities.
The prevalence of feasibility and performance compared to other metrics could be related to the lack of a ground truth of the used datasets, since they often require the number of false negatives.

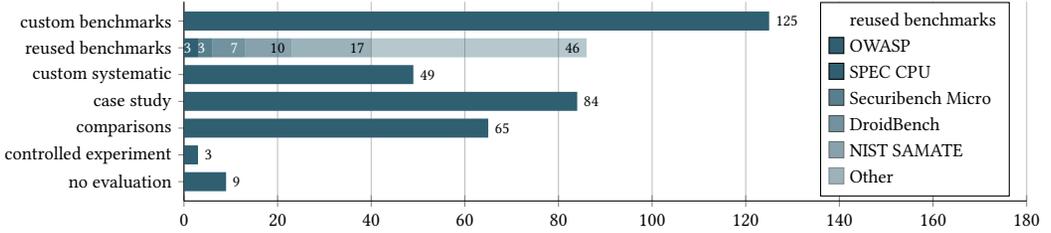
\begin{figure}[t]
    \centering
    \begin{tikzpicture}
        \scriptsize
        \begin{axis}[
            xbar stacked,
            xmin=0,
            xmax=180,
            axis lines=left,
            enlarge y limits=0.12,
            xmajorgrids=true,
            ymajorgrids=false,
            symbolic y coords={
                custom benchmarks,
                reused benchmarks,
                custom systematic,
                case study,
                comparisons,
                controlled experiment,
                no evaluation
            },
            ytick=data,
            bar width=0.25cm,
            width=.92\textwidth,
            height=4.2cm,
            y dir=reverse,
            nodes near coords,
            every node near coord/.style={font=\tiny, anchor=west},
            x axis line style={-},
            y axis line style={-},
            point meta=explicit symbolic,
            legend style={
                at={(0.98,0.999)},
                anchor=north east,
                draw=black,
                fill=white,
                font=\scriptsize,
                legend columns=1,
                legend cell align=left
            },
        ]
        \addlegendimage{empty legend}
        \addlegendentry{reused benchmarks}

        \addplot [fill={rgb:red,46; green,92; blue,110}, draw=none,]
        coordinates {
            (125,custom benchmarks)[125]
            (84,case study)[84]
            (65,comparisons)[65]
            (49,custom systematic)[49]
            (3,controlled experiment)[3]
            (9,no evaluation)[9]
            (0,reused benchmarks)[]
        };

        \addplot [fill={rgb:red,46; green,92; blue,110}, opacity=1, draw=none,
            every node near coord/.style={anchor=east, font=\tiny ,text=white,opacity=1}]
        coordinates {
            (0,custom benchmarks)[] (0,case study)[] (0,custom systematic)[]
            (0,controlled experiment)[] (0,comparisons)[](0,no evaluation)[]
            (3,reused benchmarks)[3]
        };
        \addlegendentry{OWASP}

        \addplot [fill={rgb:red,46; green,92; blue,110}, opacity=0.8, draw=none,
            every node near coord/.style={anchor=east, font=\tiny ,text=white,opacity=1}]
        coordinates {
            (0,custom benchmarks)[] (0,case study)[] (0,custom systematic)[]
            (0,controlled experiment)[](0,comparisons)[] (0,no evaluation)[]
            (3,reused benchmarks)[3]
        };
        \addlegendentry{SPEC CPU}

        \addplot [fill={rgb:red,46; green,92; blue,110}, opacity=0.68, draw=none,
            every node near coord/.style={anchor=east, font=\tiny ,text=white,opacity=1}]
        coordinates {
            (0,custom benchmarks)[] (0,case study)[] (0,custom systematic)[]
            (0,controlled experiment)[](0,comparisons)[] (0,no evaluation)[]
            (7,reused benchmarks)[7]
        };
        \addlegendentry{Securibench Micro}

        \addplot [fill={rgb:red,46; green,92; blue,110}, opacity=0.58, draw=none,
            every node near coord/.style={anchor=east, font=\tiny ,text=black,opacity=1}]
        coordinates {
            (0,custom benchmarks)[] (0,case study)[] (0,custom systematic)[]
            (0,controlled experiment)[](0,comparisons)[] (0,no evaluation)[]
            (10,reused benchmarks)[10]
        };
        \addlegendentry{DroidBench}

        \addplot [fill={rgb:red,46; green,92; blue,110}, opacity=0.48, draw=none,
            every node near coord/.style={anchor=east, font=\tiny ,text=black,opacity=1}]
        coordinates {
            (0,custom benchmarks)[] (0,case study)[] (0,custom systematic)[]
            (0,controlled experiment)[](0,comparisons)[] (0,no evaluation)[]
            (17,reused benchmarks)[17]
        };
        \addlegendentry{NIST SAMATE}

        \addplot [fill={rgb:red,46; green,92; blue,110}, opacity=0.35, draw=none,
            every node near coord/.style={anchor=east, font=\tiny ,text=black,opacity=1}]
        coordinates {
            (0,custom benchmarks)[] (0,case study)[] (0,custom systematic)[]
            (0,controlled experiment)[](0,comparisons)[] (0,no evaluation)[]
            (46,reused benchmarks)[46]
        };
        \addlegendentry{Other}

        \addplot [draw=none]
        coordinates {
            (86,reused benchmarks)[86]
        };

        \end{axis}
    \end{tikzpicture}
    \vspace{-.3cm}
    \caption{Evaluation methods for static security analyzers}
    \label{fig:benchmark_counts}
    \vspace{-.2cm}
\end{figure}

\begin{table}[b]
	\vspace{-.1cm}
	\caption{Number of papers in which each individual analyzer is used as a comparator in an evaluation}
	\label{tab:comparisons}
	\vspace{-.3cm}
    \centering
    \scriptsize
    \setlength{\tabcolsep}{6pt}
    \begin{tabular}{cl}
    \toprule
    \textbf{Number of Papers} & \textbf{Tools Used as Evaluation Comparators in Papers}
    \\ \midrule
    6 & cppcheck,\,flowdroid\\
    5 & fortify\\
    4 & flawfinder,\,rips,\,pixy\\
    3 & droidsafe,\,rats,\,iccta,\,satc,\,codeql,\,splint\\
    2 & amandroid,\,clang,\,codepro analytix,\,findbugs,\,joern,\,karonte,\,semgrep,\,cryptoguard,\,checkmarx,\,sonarqube\\
    1 & 95 other tools\\
    \bottomrule
    \end{tabular}
\end{table}

\subsection{Evaluation Methods}
We identified 5 different kinds of evaluation methods, as shown in \cref{fig:benchmark_counts}.
A total of 9 papers do not provide any kind of evaluation for their static security analysis.
We distinguish between benchmarks, which evaluate static analyzers on a larger number of systems, and case studies, which typically evaluate a small number of systems or a subset of a benchmark.
We identified 258 benchmarks, which we grouped into 3 sub-categories.

\emph{Custom benchmarks} (125 benchmarks) are most frequently used.
The systems are often selected based on the knowledge of the authors.
Although reasonings for the selections are rarely provided, some papers claim that the selected systems are either widely used, or security critical.

While most papers create custom benchmarks for their evaluation, \emph{systematically created benchmarks} (49 benchmarks) are often created by crawling data sources based on predefined criteria.
For example, many evaluations use the top X apps of several categories from the Google Play Store.
However, other platforms such as Github or Maven Central are used by applying filters such as programming languages, or number of stars or downloads.

\emph{Reused benchmarks} (86 benchmarks), e.g., benchmarks from related work, are commonly used.
These contain a collection of labeled vulnerable and non-vulnerable systems or code snippets to measure the effectiveness of static security analysis techniques on a common ground.
The most widely reused benchmarks are NIST Samate (17 papers), which contains the Juliet test suite, Droidbench (10 papers), Securibench-Micro (7 papers), the OWASP benchmark (3 papers), and SPEC CPU (3 papers).
We identified 46 further benchmarks that have only been used once or twice.

\emph{Case studies} (84 studies) evaluate analyzers on individual systems, or subsets of benchmarks.
The Linux Kernel is a common system for case studies (10 studies).
In 32 papers, case studies are combined with other evaluation methods, often highlighting findings from larger benchmarks.

\emph{Comparisons} (65 comparisons) to other static security analyzers are often performed to evaluate the effectiveness of the proposed analyzer in relation to existing ones.
The most frequently used tools for comparisons, listed in \cref{tab:comparisons}, are cppcheck and flowdroid, which are used in 6 papers each, followed by fortify (5 papers), flawfinder, rips, and pixy (4 papers each).
On average, comparisons are performed with 2.48 other tools.
When comparing the reuse of benchmarks and comparisons, we found that a few benchmarks are more frequently reused than individual tools.

\emph{Controlled experiments} (3 papers) are performed to assess the usability of static security analyzers.

\subsection{Evaluation Size}

\begin{figure}[t]
\centering
\begin{tikzpicture}[
  every axis/.style={
    ybar stacked,
    ymin=0,
    ymax=60,
    x tick label style={font=\scriptsize},
    symbolic x coords={0,1,2-4,5-9,10-24,25-49,50-99,100-249,250-499,500-999,1000+,z},
    xticklabels={,,1,2-4,5-9,10-24,25-49,50-99,100-249,250-499,500-999,1000+,}, 
    bar width=12pt,
    width=\textwidth,
    height=3.8cm,
    ytick distance=10,
    ymajorgrids=true,
    axis y line=left,
    axis x line=bottom,
    x axis line style={-},
  }
]

        \scriptsize
\begin{axis}[bar shift=-6pt,
          nodes near coords,
          nodes near coords align={vertical},
          every node near coord/.append style={
              font=\tiny,
              xshift=-6pt,
              yshift=-7pt,
              text=black,
              opacity=1
          },
          nodes near coords={\pgfplotspointmeta},
          point meta=explicit symbolic,
          legend style={
              at={(.86,.95)},
              anchor=north east,
              draw=black,
              fill=white,
              font=\scriptsize,
              legend columns=2,
              legend cell align=left
          },
            legend image post style={
                draw=black,
                line width=0.1pt
            },
                legend image code/.code={
                \path[draw=#1, fill=#1] (-0.07cm,-0.07cm) rectangle (0.07cm,0.07cm);
            }
        x axis line style={-},
        y axis line style={-},
]
\addplot+[fill={rgb,255:red,242;green,142;blue,43}, draw=none,
          every node near coord/.append style={
              font=\tiny,
              text=black,
              yshift=-.5pt,
              opacity=1
          },] coordinates
    {(0,0) [] (1,44) [44] (2-4,25) [25] (5-9,5) [5] (10-24,0) [] (25-49,0) [] (50-99,0) [] (100-249,0) [] (250-499,0) [] (500-999,0) [] (1000+,0) [] (z,0) []};
\addlegendentry{Case Study - Projects}

\addplot+[fill={rgb,255:red,242;green,142;blue,43}, opacity=0.35, draw=none,
          every node near coord/.append style={
              font=\tiny,
              yshift=5pt,
              text=black,
              opacity=1
          },] coordinates
    {(0,0) [] (1,4) [4] (2-4,1) [1] (5-9,2) [2] (10-24,3) [3] (25-49,0) [] (50-99,0) [] (100-249,0) [] (250-499,0) [] (500-999,0) [] (1000+,0) [] (z,0) []};
\addlegendentry{Case Study - Examples}

\addlegendimage{ybar, fill={rgb:red,46; green,92; blue,110}, draw=black}
\addlegendentry{Benchmark - Projects}
\addlegendimage{ybar, fill={rgb:red,46; green,92; blue,110}, opacity=0.35} 
\addlegendentry{Benchmark - Examples}

\end{axis}

\begin{axis}[bar shift=6pt, hide axis,
          nodes near coords,
          nodes near coords align={vertical},
          every node near coord/.append style={
              font=\tiny,
              xshift=6pt,
              yshift=-7pt,
              text=black,
              opacity=1
          },
          nodes near coords={\pgfplotspointmeta},
          point meta=explicit symbolic,
]
\addplot+[fill={rgb:red,46; green,92; blue,110}, draw=none,
          every node near coord/.append style={
              font=\tiny,
              yshift=-.5pt,
              text=white,
              opacity=1
          }] coordinates
    {(0,0) [] (1,0) [] (2-4,19) [19] (5-9,46) [46] (10-24,37) [37] (25-49,14) [14] (50-99,15) [15] (100-249,15) [15] (250-499,7) [7] (500-999,11) [11] (1000+,37) [37] (z,0) []};

\addplot+[fill={rgb:red,46; green,92; blue,110}, draw=none, opacity=0.35,
          every node near coord/.append style={
              font=\tiny,
              yshift=5pt,
              text=black,
              opacity=1
          },] coordinates
    {(0,0) [] (1,0) [] (2-4,2) [2] (5-9,5) [5] (10-24,7) [7] (25-49,9) [9] (50-99,7) [7] (100-249,6) [6] (250-499,4) [4] (500-999,0) [] (1000+,18) [18] (z,0) []};
\end{axis}

\end{tikzpicture}
\vspace{-.3cm}
\caption{Empirical evaluation counts by study type and system size}
\label{fig:size_counts}
\vspace{-.2cm}
\end{figure}
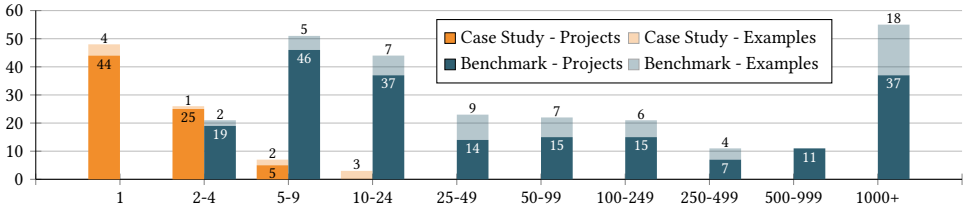

We analyzed the size of benchmarks and case studies used to evaluate static security analyzers, shown in \cref{fig:size_counts}.
We differentiate between full projects, which contain all source code of a software system, and examples, which are artificial code snippets to showcase specific vulnerabilities.

Most benchmarks with projects are rather small in size, ranging between 5 and 9 projects (44 benchmarks), or 10 to 24 projects (37 benchmarks), and are often custom created ones.
When using a systematically created benchmark, they are often larger than 1000 projects (36 benchmarks).
Such benchmarks are often created by crawling repositories from platforms such as Github, Maven Central, or the Google Play Store, and either contain self-contained systems or libraries.

Benchmarks often exceed 1000 examples (18 benchmarks) when reusing benchmarks such as NIST Samate, including the Juliet test suite, covering many different CWEs.

Not surprisingly, most case studies include single projects or examples (46 studies) to evaluate a static security analyzer on one system, or to highlight findings from benchmarks.
While a few case studies include to demonstrate effectiveness on 10-24 examples (3 studies), they rarely rely on artificial benchmarks and rather use real-world projects such as the Linux kernel.

\subsection{Formal Proofs}

\begin{figure}[t]
    \centering
    \begin{tikzpicture}
        \scriptsize
        \begin{axis}[
            xbar,
            xmin=0,
            xmax=10,
            xlabel=,
            ylabel=,
            axis lines=left,
            enlarge y limits=0.15,
            xmin=0,
            xmajorgrids=true,
            ymajorgrids=false,
            symbolic y coords={
                soundness,
                correctness,
                algorithm properties,
                complexity,
                sufficiency,
            },
            ytick=data,
            bar width=0.25cm,
            width=.9\textwidth,
            height=3cm,
            y dir=reverse,
            nodes near coords,
            every node near coord/.style={font=\tiny, anchor=east},
            x axis line style={-},
            y axis line style={-},
            point meta=explicit symbolic,
        ]
        \addplot [fill={rgb:red,46; green,92; blue,110}, draw=none,
            every node near coord/.style={anchor=west, font=\tiny ,text=black,opacity=1}] coordinates {
            (8,soundness)[8]
            (6,correctness)[6]
            (5,algorithm properties)[5]
            (5,complexity)[5]
            (1,sufficiency)[1]
        };
        \end{axis}
    \end{tikzpicture}
    \vspace{-.4cm}
    \caption{Provided formal proofs for static security analyzers}
    \label{fig:proofs_counts}
    \vspace{-.3cm}
\end{figure}
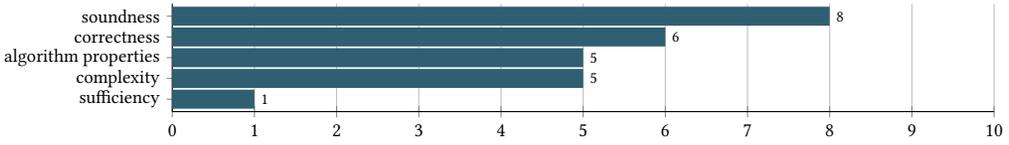

\looseness=-1
Only 19 papers support their analysis with formal proofs, listed in \cref{fig:proofs_counts}.
\emph{Soundness} (8 papers) shows that the analysis does not produce any false positives or negatives, while \emph{correctness} (6 papers) proves that the analysis correctly identifies vulnerabilities according to its specifications.
\emph{Complexity} (5 papers) describes the computational complexity of the analysis in terms of time or space.
\emph{Sufficiency} (1 paper) proves that if a condition is met, a specific output is guaranteed.
While few papers provide proofs for their analyses, their focus is often set on formalizing static analysis techniques and formal verification.
Their evaluation is often limited to small custom code snippets. 


\quotebox{Evaluation (RQ4)}{Empirical evaluations typically use custom benchmarks, case studies, systematically created benchmarks, or previously established benchmarks such as NIST Samate or Droidbench. The majority of benchmarks contain only a small numbers of projects, though systematically created ones often include thousands of systems. Feasibility and performance are the most frequently used evaluation metrics. Only a few papers provide formal proofs such as for soundness or correctness.}{5cm}

\section{Limitations of Static Security Analyzers (RQ5)}
To understand the limitations of static security analyzers, we investigated explicitly mentioned ones, the assumptions made in attacker models, as well as the exploitability of detected vulnerabilities.

\subsection{Limitations}

\begin{figure}[t]
    \centering
    \begin{tikzpicture}
        \scriptsize
        \begin{axis}[
            xbar,
            xmin=0,
            xmax=55,
            xlabel=,
            ylabel=,
            axis lines=left,
            enlarge y limits=0.07,
            xmin=0,
            xmajorgrids=true,
            ymajorgrids=false,
            symbolic y coords={
                approximations,
                scope,
                missing language features,
                dynamic features,
                multi-threading,
                overhead,
                performance,
                native code,
                presentation,
                missing proofs,
                exploitability
            },
            ytick=data,
            bar width=0.25cm,
            width=.9\textwidth,
            height=4.8cm,
            y dir=reverse,
            nodes near coords,
            every node near coord/.style={font=\tiny, anchor=east},
            x axis line style={-},
            y axis line style={-},
            point meta=explicit symbolic,
        ]
        \addplot [fill={rgb:red,46; green,92; blue,110}, draw=none,
            every node near coord/.style={anchor=west, font=\tiny ,text=black,opacity=1}] coordinates {
            (52,approximations)[52]
            (30,scope)[30]
            (29,missing language features)[29]
            (21,dynamic features)[21]
            (8,multi-threading)[8]
            (7,overhead)[7]
            (6,performance)[6]
            (5,native code)[5]
            (4,presentation)[4]
            (3,missing proofs)[3]
            (2,exploitability)[2]
        };
        \end{axis}
    \end{tikzpicture}
    \vspace{-.8cm}
    \caption{Discussed limitations of static security analyzers}
    \label{fig:limitation_counts}
    \vspace{-.4cm}
\end{figure}
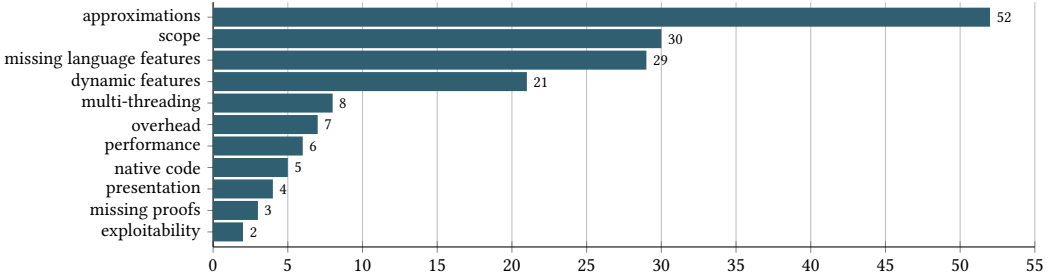

A total of 99 papers explicitly discuss limitations of their static security analyzers.
We grouped similar limitations into 11 categories as shown in \cref{fig:limitation_counts}.

\emph{Approximations} (52 papers) can lead to false positives or negatives, as the analysis may not accurately capture all possible program behaviors or states.
For example, a taint analysis may taint a whole data structure such arrays once a tainted variable is stored in it, leading to false positives.

Many static security analyzers are limited to specific programming languages or domains, and may not be applicable outside this \emph{scope} (30 papers) without significant modifications or adaptations.

Even within the scope, static security analyzers may \emph{miss language features} (29 papers), leading to missed vulnerabilities or incomplete analyses, as certain code constructs may not be analyzed.

Since static security analyzers only analyze the source code without executing it, they are unable to handle \emph{dynamic features} (21 papers) of programming languages such as dynamic code generation, reflection, or runtime type information, leading to potentially missed vulnerabilities.

Further limitations include the inability to handle \emph{multi-threading} (8 papers), \emph{overhead} in terms of maintenance of updating rules or vulnerability databases (7 papers), \emph{performance} issues such as long analysis times (6 papers), or the inability to analyze \emph{native code} (5 papers).
Finally, individual papers discuss limitations in \emph{presentation} of results (4 papers), missing formal proofs of the analysis (3 papers), or the \emph{non-exploitability} of detected vulnerabilities (2 papers).

A shockingly large number of papers (147 papers) do not discuss any limitations at all.
Although some limitations such as false positives are apparent from the results, they are not explicitly discussed.
As a result, the root causes of the analyzer's shortcomings remain unclear.

\subsection{Attacker Models}
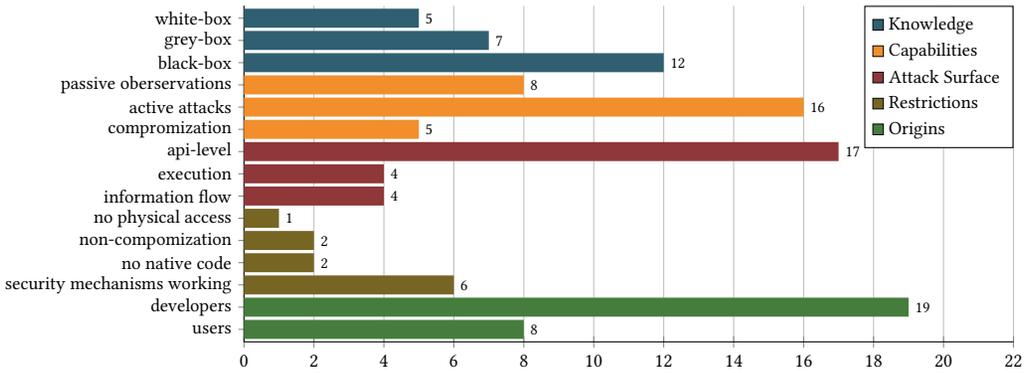
\begin{figure}[t]
    \centering
    \begin{tikzpicture}
    \scriptsize
    \begin{axis}[
        xbar,
        xmin=0,
        xmax=22,
        axis lines=left,
        enlarge y limits=0.04,
        xmajorgrids=true,
        ymajorgrids=false,
        symbolic y coords={
            white-box,
            grey-box,
            black-box,
            passive oberservations,
            active attacks,
            compromization,
            api-level,
            execution,
            information flow,
            no physical access,
            non-compomization,
            no native code,
            security mechanisms working,
            developers,
            users
        },
        ytick=data,
        yticklabels={
            white-box,
            grey-box,
            black-box,
            passive oberservations,
            active attacks,
            compromization,
            api-level,
            execution,
            information flow,
            no physical access,
            non-compomization,
            no native code,
            security mechanisms working,
            developers,
            users
        },
        y dir=reverse,
        bar width=0.25cm,
        width=.85\textwidth,
        height=6cm,
        nodes near coords,
        x axis line style={-},
        y axis line style={-},
        point meta=explicit symbolic,
        every node near coord/.style={anchor=west,font=\tiny,text=black,opacity=1},
        every axis plot/.append style={
          bar shift=0pt
        },
        legend style={
            at={(1,1)},
            anchor=north east,
            draw=black,
            fill=white,
            font=\scriptsize,
            legend columns=1,
            legend cell align=left
        },
            legend image post style={
            draw=black,
            line width=0.1pt
        },
            legend image code/.code={
            \path[draw=#1, fill=#1] (-0.07cm,-0.07cm) rectangle (0.07cm,0.07cm);
        }
    ]

    \addplot [fill={rgb:red,46;green,92;blue,110}, draw=none] coordinates {
        (5,white-box)[5]
        (7,grey-box)[7]
        (12,black-box)[12]
        (0,passive oberservations)[]
        (0,active attacks)[]
        (0,compromization)[]
        (0,api-level)[]
        (0,execution)[]
        (0,information flow)[]
        (0,no physical access)[]
        (0,non-compomization)[]
        (0,no native code)[]
        (0,security mechanisms working)[]
        (0,developers)[]
        (0,users)[]
    };
    \addlegendentry{Knowledge}

    \addplot [fill={rgb,255:red,242;green,142;blue,43}, draw=none] coordinates {
        (0,white-box)[]
        (0,grey-box)[]
        (0,black-box)[]
        (8,passive oberservations)[8]
        (16,active attacks)[16]
        (5,compromization)[5]
        (0,api-level)[]
        (0,execution)[]
        (0,information flow)[]
        (0,no physical access)[]
        (0,non-compomization)[]
        (0,no native code)[]
        (0,security mechanisms working)[]
        (0,developers)[]
        (0,users)[]
    };
    \addlegendentry{Capabilities}

    \addplot [fill={rgb:red,225;green,87;blue,89}, draw=none] coordinates {
        (0,white-box)[]
        (0,grey-box)[]
        (0,black-box)[]
        (0,passive oberservations)[]
        (0,active attacks)[]
        (0,compromization)[]
        (17,api-level)[17]
        (4,execution)[4]
        (4,information flow)[4]
        (0,no physical access)[]
        (0,non-compomization)[]
        (0,no native code)[]
        (0,security mechanisms working)[]
        (0,developers)[]
        (0,users)[]
    };
    \addlegendentry{Attack Surface}

    \addplot [fill={rgb:red,237;green,201;blue,72}, draw=none] coordinates {
        (0,white-box)[]
        (0,grey-box)[]
        (0,black-box)[]
        (0,passive oberservations)[]
        (0,active attacks)[]
        (0,compromization)[]
        (0,api-level)[]
        (0,execution)[]
        (0,information flow)[]
        (1,no physical access)[1]
        (2,non-compomization)[2]
        (2,no native code)[2]
        (6,security mechanisms working)[6]
        (0,developers)[]
        (0,users)[]
    };
    \addlegendentry{Restrictions}

    \addplot [fill={rgb:red,89;green,161;blue,79}, draw=none] coordinates {
        (0,white-box)[]
        (0,grey-box)[]
        (0,black-box)[]
        (0,passive oberservations)[]
        (0,active attacks)[]
        (0,compromization)[]
        (0,api-level)[]
        (0,execution)[]
        (0,information flow)[]
        (0,no physical access)[]
        (0,non-compomization)[]
        (0,no native code)[]
        (0,security mechanisms working)[]
        (19,developers)[19]
        (8,users)[8]
    };
    \addlegendentry{Origins}

    \end{axis}
    \end{tikzpicture}
    \vspace{-.35cm}
    \caption{Attacker model dimensions considered by static security analyzers}
    \label{fig:attacker_model_counts}
    \vspace{-.5cm}
\end{figure}

Attacker models are described in 28 papers, primarily from cybersecurity conferences or journals.
We found assumptions made on the required \emph{knowledge} to exploit a vulnerability, \emph{capabilities} that attackers have, the \emph{attack surface}, certain \emph{restrictions} that apply, and the \emph{origin} of the attack.

Most attacker models assume \emph{black-box} (i.e., none) knowledge (12 papers) or \emph{grey-box} (i.e., some) knowledge (7 papers) about a software system, while only few assume \emph{white-box} knowledge (5 papers), which requires full access to the source code.
These are reasonable assumptions, considering attackers typically have no access to the source code of proprietary software systems, but may make certain assumptions (e.g., the presence of a specific dependency) or have access to some parts of the system (e.g., are able to read certain variables).
Still, attackers may have access to firmware of IoT devices or browser extensions, allowing them to analyze their source code. 

The majority of attacker models describe \emph{active attacks} (16 papers) on a system such as injection attacks.
Still, many assume the attacker to be a \emph{passive observer} (8 papers), who only analyzes the output of a system for e.g., side channel attacks. 
Five papers assume the attacker has \emph{compromized} a whole component of the system or the environment it is deployed in.

Since most attacker models assume active attacks, they usually assume the \emph{system's API} to be the attack surface (17 papers).
For example, some attacker models assume developers to make mistakes when using cryptographic APIs, and write vulnerable code.
In other instances, only the \emph{execution} (4 papers) of the system is observed to e.g., measure differences in computation time for side channel attacks.
Observing or capturing the content of \emph{information flow} is assumed in 4 papers.

Some attacker models impose restrictions, assuming that a system's \emph{security mechanisms are working} and cannot be bypassed (6 papers).
For example, attackers are assumed unable to bypass the Android permission model, or access control checks.
Other models exclude \emph{native code} (2 papers), \emph{compromization} of system components (2 papers), or \emph{physical access} (1 paper) from their scope.

Finally, most attacker models assume the origin or cause of attacks to be a \emph{developer} (19 papers) or user who is able to write malicious code.
The remaining attacker models assume that \emph{users} (8 papers) are not able to do so.
Therefore, the two common patterns are (i) developers proficient in writing malicious code to perform active attacks on software APIs, and (ii) users with no coding skills to passively observe the execution based on user input, or implicit information flow.

\subsection{Exploitability}

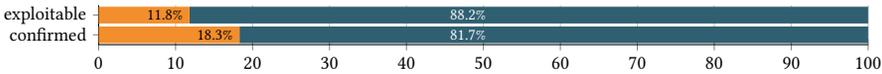
\begin{figure}[t]
    \centering
    \begin{tikzpicture}
        \scriptsize
        \begin{axis}[
            xbar stacked,
            xmin=0,
            xmax=100,
            xlabel=,
            ylabel=,
            axis lines=left,
            enlarge y limits=0.5,
            xmin=0,
            xmajorgrids=true,
            ymajorgrids=false,
            symbolic y coords={
                exploitable,
                confirmed
            },
            ytick=data,
            bar width=0.22cm,
            width=.85\textwidth,
            height=2.1cm,
            y dir=reverse,
            nodes near coords,
            every node near coord/.style={font=\tiny, anchor=east},
            x axis line style={-},
            y axis line style={-},
            point meta=explicit symbolic,
        ]
        \addplot [fill={rgb,255:red,242;green,142;blue,43}, draw=none,
            every node near coord/.style={anchor=east, font=\tiny ,text=black,opacity=1}] coordinates {
            (11.8,exploitable)[11.8\%]
            (18.3,confirmed)[18.3\%]
        };
        \addplot [fill={rgb:red,46; green,92; blue,110}, draw=none,
            every node near coord/.style={anchor=east, font=\tiny ,text=black,opacity=1, text=white, xshift=-5cm}] coordinates {
            (88.2,exploitable)[88.2\%]
            (81.7,confirmed)[81.7\%]
        };
        \end{axis}
    \end{tikzpicture}
	\vspace{-.45cm}
    \caption{Percentage of papers reporting to have detected exploitable or confirmed vulnerabilities}
    \label{fig:ex_con_counts}
    \vspace{-.22cm}
\end{figure}

For each paper, we tracked whether the authors reported whether the detected vulnerabilities were exploitable by an attacker, or whether they were confirmed by developers of the evaluated systems (\cref{fig:ex_con_counts}).
While most papers we investigated provide an evaluation on the effectiveness of static security analyzers, only 11.8\% of papers claimed to have found vulnerabilities that are exploitable by an attacker.
For example, an analyzer may report a potential SQL injection due to unsanitized input, yet, in practice, the input may be constrained by unreachable execution paths.
Although this could be seen as a vulnerability, it has no practical impact since it cannot be exploited.

In addition, 18.3\% of papers report vulnerabilities confirmed by developers of the evaluation systems, or that have been assigned a CVE.
In these cases, it is typically reported that the developers also fixed the vulnerability based on the report of the static security analyzer, which shows practical impact.
Most notably, the Linux Kernel has been frequently reported by the papers to have vulnerabilities that have been fixed by developers, since it is a widely used and maintained system.

\quotebox{Limitations (RQ5)}{Most static security analysis papers do not explicitly report any limitations, making it difficult to assess their shortcomings. The most common limitations are approximations leading to false positives or negatives, limited language- or domain-specific scope, missing code elements, and inability to handle dynamic language features. Attacker models are reported in few papers, typically assuming active attacks by developers on software APIs or passive observation by users. Only a minority of analyzers report exploitable vulnerabilities or vulnerabilities confirmed by developers.}{5cm}

%% file: tables/analyzers-longtable.tex
\vspace{4pt}
\setlength{\tabcolsep}{1.5pt}
\scriptsize
\begin{tabularx}{\textwidth}{llllXl}
\caption{
\centering
Static security analyzers included in our survey, and selected characteristics\\
\smallskip
\scriptsize
\textit{df\,=\,dataflow, cf\,=\,control flow, s\,=\,structural code elements, dfa\,=\,dataflow analysis, cfa\,=\,control flow analysis, pm\,=\,pattern matching, mc\,=\,model checking, se\,=\,symbolic execution, smt\,=\,SMT solving, ps\,=\,program slicing, ra\,=\,reference analysis, va\,=\,value analysis, da\,=\,dependence analysis, ta\,=\,type analysis, la\,=\,linker analysis, cla\,=\,concurrency/lock analysis, sta\,=\,string analysis, ba\,=\,bounds analysis.}\vspace{-8pt}
}
\\
\toprule
\textbf{Analyzer} & \textbf{Year} & \textbf{Language} & \textbf{Analyzes} & \textbf{CWE Comprehensive Categorization} & \textbf{Techniques} \\
\midrule
\endfirsthead

\toprule
\textbf{Analyzer} & \textbf{Year} & \textbf{Language} & \textbf{Analyzes} & \textbf{CWE Comprehensive Categorization} & \textbf{Techniques} \\
\midrule
\endhead

\midrule
\multicolumn{6}{r}{\emph{Continued on next page}} \\
\endfoot

\bottomrule
\endlastfoot
Wagner et al.\,\cite{Wagner2000A} & 2000 & c/c++ & df & 1399, 1406, 1407, 1416 & smt \\
Splint\,\cite{Larochelle2001Statically} & 2001 & c/c++ & df & 1399, 1406, 1407, 1416 & dfa, ra, va \\
Eau Claire\,\cite{Chess2002} & 2002 & c/c++ & s & 1399, 1401, 1406, 1407, 1410, 1416 & smt \\
MOPS\,\cite{Chen2002MOPS} & 2002 & c/c++ & cf & 1410 & mc \\
ARCHER\,\cite{Xie2003ARCHER} & 2003 & c/c++ & cf & 1399, 1406, 1407, 1416 & se \\
Avvenuti et al.\,\cite{Avvenuti2003Java} & 2003 & java & df & 1403, 1416, 1417 & dfa \\
CodeSurfer\,\cite{Ganapathy2003Buffer} & 2003 & c/c++ & df & 1399, 1406, 1407, 1416 & dfa, ps, ra, smt \\
CSSV\,\cite{Dor2003CSSV} & 2003 & c/c++ & cf & 1399, 1406, 1407, 1416 & cfa, ra, va \\
MECA\,\cite{Yang2003MECA} & 2003 & c/c++ & df & 1405, 1412 & dfa \\
Benjamin et al.\,\cite{Benjamin2005Finding} & 2005 & java & df & 1398, 1403, 1404, 1406, 1407, 1409, 1413, 1416 & dfa, ra \\
Ceesay et al.\,\cite{Ceesay2006Using} & 2006 & c/c++ & df & 1406, 1407, 1408 & dfa \\
Cova et al.\,\cite{Cova2006Static} & 2006 & c/c++ & df & 1406, 1407 & dfa, ps, ra, se \\
Heine et al.\,\cite{Heine2006Static} & 2006 & c/c++ & cf & 1416 & ra \\
Xie et al.\,\cite{Xie2006Static} & 2006 & php & cf & 1407, 1409 & cfa, pm, se \\
Liu et al.\,\cite{Liu2007Bytecode} & 2007 & java & cf & 1396, 1406, 1407 & ta \\
Pistoia et al.\,\cite{Pistoia2007When} & 2007 & java & df & 1396 & dfa \\
SAFELI\,\cite{Fu2007A} & 2007 & c\# & cf & 1407, 1409 & se \\
SAFES\,\cite{Shi2007SAFES} & 2007 & java & s & 1396, 1403, 1416, 1417 & ra \\
Sarkar et al.\,\cite{Sarkar2007Flowinsensitive} & 2007 & c/c++ & s & 1406, 1407, 1408 & pm \\
Wassermann et al.\,\cite{Wassermann2007Sound} & 2007 & php & df & 1407, 1409 & dfa \\
Almaliotis et al.\,\cite{Almaliotis2008Static} & 2008 & java & cf, df & 1396, 1405 & cfa, dfa, ta \\
Beckman et al.\,\cite{Beckman2008Verifying} & 2008 & java & df & 1401, 1410 & dfa, pm \\
FADO\,\cite{Vujosevic-Janicic2008Ensuring} & 2008 & c/c++ & df & 1399, 1406, 1407, 1416 & dfa, smt \\
Liang et al.\,\cite{Liang2008Automatic} & 2008 & c/c++ & df & 1406, 1407, 1408 & ta \\
Liu et al.\,\cite{Liu2008Static} & 2008 & java & df & 1403, 1416, 1417 & dfa, ra \\
Scholz et al.\,\cite{Scholz2008Userinput} & 2008 & c/c++ & df & 1406, 1407 & cfa, dfa \\
Tlili et al.\,\cite{Tlili2008A} & 2008 & c/c++ & df & 1399, 1406, 1407, 1416 & ta \\
Wassermann et al.\,\cite{Wassermann2008Static} & 2008 & php & df & 1407, 1409 & dfa, sta \\
Yu et al.\,\cite{Yu2008Symbolic} & 2008 & php & s & 1406, 1407 & va \\
A3\,\cite{Geay2009Modular} & 2009 & java & df & 1396 & la, ps, sta \\
Dasgupta et al.\,\cite{Dasgupta2009A} & 2009 & c\# & df & 1407, 1409 & dfa \\
DUVP\,\cite{Shushen2009Static} & 2009 & php & s & 1406, 1407 & pm \\
Li et al.\,\cite{Li2009Finding} & 2009 & c/c++ & df & 1405 & dfa \\
Liang et al.\,\cite{Liang2009Static} & 2009 & c/c++ & df & 1399, 1406, 1407, 1416 & dfa \\
Liu et al.\,\cite{Liu2009Practical} & 2009 & java & df & 1403, 1416, 1417 & dfa, ra \\
Svace\,\cite{Nesov2009Automatically} & 2009 & c/c++ & df & 1399, 1401, 1403, 1405, 1406, 1407, 1410, 1412, 1415, 1416 & dfa, ra, va \\
CSV\,\cite{Zheng2010Research} & 2010 & c/c++ & s & 1399, 1406, 1407, 1412, 1416 & pm \\
Li et al.\,\cite{Li2010Practical} & 2010 & c/c++ & cf & 1399, 1406, 1407, 1416 & cfa, se \\
Liu et al.\,\cite{Liu2010Static} & 2010 & java & df & 1406, 1407 & dfa, ra \\
PDVDS\,\cite{Cheng2010PDVDS} & 2010 & c/c++ & df & 1399, 1403, 1406, 1407, 1408, 1409, 1416 & dfa \\
Pixy (extension)\,\cite{Jovanovic2010Static} & 2010 & php & df & 1407, 1409 & cfa, dfa, ra, va \\
STAC\,\cite{Cearǎ2010Taint} & 2010 & c/c++ & cf & 1399, 1406, 1407, 1416 & cfa, ra \\
Tasa\,\cite{Huang2010Vulnerabilities} & 2010 & c\# & df & 1407, 1409 & dfa, ra \\
VD-PTA\,\cite{Akbari2010Vulnerability} & 2010 & c/c++ & s & 1399, 1401, 1406, 1407, 1410, 1416 & pm \\
Rolecast\,\cite{Son2011Rolecast} & 2011 & java & df & 1396 & da, dfa, pm \\
SAFERPHP\,\cite{Son2011SAFERPHP} & 2011 & php & df & 1396, 1407, 1409, 1416 & cfa, dfa, ra, se \\
Side\,Channel\,Finder\,\cite{Lux2011A} & 2011 & java & cf, df & 1403, 1416, 1417 & dfa \\
Wang et al.\,\cite{Wang2011Program} & 2011 & php & df & 1407, 1409 & dfa, ps \\
Almorsy et al.\,\cite{Almorsy2012Supporting} & 2012 & c\#,\,vb.net,\,c/c++ & cf, df & 1396, 1403, 1407, 1409, 1411, 1413, 1416, 1417 & cfa, dfa, pm \\
AndroidLeaks\,\cite{Gibler2012AndroidLeaks} & 2012 & java & df & 1403, 1416, 1417 & dfa \\
Berger et al.\,\cite{Berger2012An} & 2012 & java & df & 1403, 1416, 1417 & dfa \\
Codeminer\,\cite{Agosta2012Automated} & 2012 & php & df & 1407, 1409 & dfa, se, sta \\
DroidChecker\,\cite{Chan2012DroidChecker} & 2012 & java & df & 1396, 1405 & dfa \\
JCSI\,\cite{Avvenuti2012JCSI} & 2012 & java & cf, df & 1403, 1416, 1417 & dfa \\
Kint\,\cite{Wang2012Improving} & 2012 & c/c++ & cf, df & 1406, 1407, 1408 & dfa, smt \\
Lu et al.\,\cite{Lu2012Modelbased} & 2012 & java & df & 1396 & dfa, smt \\
Rawat et al.\,\cite{Rawat2012Finding} & 2012 & c/c++ & s & 1399, 1406, 1407, 1416 & cfa, la \\
Sifa\,\cite{Mills2012ToolSupported} & 2012 & c/c++ & df & 1403, 1416, 1417 & dfa \\
Vedala et al.\,\cite{Vedala2012Automatic} & 2012 & c/c++ & s & 1399, 1403, 1406, 1407, 1416 & pm \\
Andromeda\,\cite{Tripp2013Andromeda} & 2013 & java, c\#, js & df & 1403, 1416, 1417 & dfa, ra \\
Chucky\,\cite{Yamaguchi2013Chucky} & 2013 & c/c++ & df & 1406, 1407 & dfa, pm \\
Fehnker et al.\,\cite{Fehnker2013Model} & 2013 & c/c++ & cf & 1405, 1412, 1416 & cfa, mc, se, smt \\
FixMeUp\,\cite{Son2013Fix} & 2013 & php & cf & 1396 & ps, va \\
HAVOC-LITE\,\cite{Vanegue2013Towards} & 2013 & c/c++ & df & 1396, 1399, 1402, 1403, 1408, 1413, 1416, 1417 & cfa, dfa, smt \\
Kirrage et al.\,\cite{Kirrage2013Static} & 2013 & - & s & 1397 & pm \\
STOP\,\cite{Goichon2013Static} & 2013 & java & df & 1396, 1415, 1416 & dfa \\
Zheng et al.\,\cite{Zheng2013Path} & 2013 & php & cf & 1407, 1409, 1416 & cfa, va \\
Amandroid\,\cite{Wei2014Amandroid} & 2014 & java & df & 1396, 1403, 1407, 1409, 1416, 1417 & cfa, da, dfa, pm \\
Antoshina et al.\,\cite{Antoshina2014A} & 2014 & while & df & 1403, 1416, 1417 & dfa \\
Baojiang et al.\,\cite{Baojiang2014Reverse} & 2014 & java & df & 1407, 1409 & dfa \\
Bartel et al.\,\cite{Bartel2014Static} & 2014 & java & cf & 1396, 1412, 1418 & va \\
Dahse et al.\,\cite{Dahse2014Static} & 2014 & php & cf & 1407, 1409 & dfa \\
FlowDroid\,\cite{Arzt2014FLOWDROID} & 2014 & java & df & 1403, 1416, 1417 & dfa \\
FUSE\,\cite{Ravitch2014Multiapp} & 2014 & java & df & 1403, 1416, 1417 & dfa, sta \\
GreenArrays\,\cite{Nazaré2014Validation} & 2014 & c/c++ & cf, df & 1399, 1406, 1407, 1408, 1416 & cfa, dfa \\
Rimsa et al.\,\cite{Rimsa2014Efficient} & 2014 & php & df & 1407, 1409, 1416 & dfa \\
RIPS\,\cite{Dahse2014Simulation} & 2014 & php & df & 1406, 1407 & cfa, dfa, va \\
SAFEWAPI\,\cite{Bae2014SAFEWAPI} & 2014 & js & df & 1396 & ta \\
Sun et al.\,\cite{Sun2014Detecting} & 2014 & php & cf, df & none & dfa, se \\
Tlili et al.\,\cite{Tlili2014Scalable} & 2014 & c/c++ & cf & 1410 & cfa, ra \\
WARlord\,\cite{Møller2014Automated} & 2014 & java & df & 1396, 1403, 1416 & dfa \\
Yamaguchi et al.\,\cite{Yamaguchi2014Modeling} & 2014 & c/c++ & cf, df & 1396, 1399, 1401, 1403, 1405, 1406, 1407, 1408, 1410, 1412, 1415, 1416, 1417, 1418 & cfa, dfa \\
AAPL\,\cite{Lu2015Checking} & 2015 & java & df & 1403, 1416, 1417 & dfa \\
CacheAudit\,\cite{Doychev2015CacheAudit} & 2015 & c/c++ & df & 1403, 1416, 1417 & dfa \\
Julia fs/fi\,\cite{Ernst2015Boolean} & 2015 & java & df & 1406, 1407, 1409, 1415, 1416 & dfa \\
MorphDroid\,\cite{Ferrara2015MORPHDROID} & 2015 & java & df & 1403, 1416, 1417 & dfa \\
Muntean et al.\,\cite{Muntean2015SMTconstrained} & 2015 & c/c++ & cf & 1406, 1407, 1408, 1416 & ps, se, smt \\
phpSAFE\,\cite{Nunes2015PhpSAFE} & 2015 & php & df & 1407, 1409 & dfa \\
SADroid\,\cite{Han2015Systematic} & 2015 & java & df & 1396, 1403, 1416, 1417 & dfa, pm \\
VulHunter\,\cite{Qian2015VulHunter} & 2015 & java & df & 1403, 1406, 1407, 1410, 1416, 1417 & ps, smt \\
Arroyo et al.\,\cite{Arroyo2016An} & 2016 & c/c++ & df & 1399, 1403, 1406, 1407, 1409, 1416, 1417 & dfa, se \\
CrossFire\,\cite{Buyukkayhan2016CrossFire} & 2016 & js & df & 1416 & dfa \\
Entroine\,\cite{Stergiopoulos2016Execution} & 2016 & java & s & 1398, 1406, 1407, 1409 & cfa, pm \\
ESVD\,\cite{Sampaio2016Exploring} & 2016 & java & df & 1398, 1403, 1404, 1406, 1407, 1409, 1413, 1415, 1416 & dfa \\
HornDroid\,\cite{Calzavara2016HornDroid} & 2016 & java & df & 1403, 1416, 1417 & dfa, smt \\
HybriDroid\,\cite{Lee2016HybriDroid} & 2016 & java, js & df & 1403, 1416, 1417 & dfa, va \\
JS-QL\,\cite{Nicolay2016Static} & 2016 & js & df & 1396, 1402, 1403, 1407, 1409, 1412, 1413, 1415, 1416 & mc \\
JSPChecker\,\cite{Steinhauser2016JSPChecker} & 2016 & java & df & 1407, 1409 & dfa, va \\
Julia\,\cite{Spoto2016The} & 2016 & java & cf, df & 1407, 1409 & cfa, cla, dfa, ra, smt, va \\
Lester et al.\,\cite{Lester2016Information} & 2016 & js & df & 1403, 1416, 1417 & dfa \\
MLSA\,\cite{Liang2016MLSA} & 2016 & c/c++, fortran & cf & 1408, 1412 & cfa, ra, se, smt \\
Mélange\,\cite{Shastry2016Towards} & 2016 & c/c++ & cf & 1416 & cfa, se \\
R-Droid\,\cite{Backes2016RDroid} & 2016 & java & df & 1396, 1403, 1416, 1417 & cfa, da, dfa, ps, va \\
Stack\,\cite{Wang2016A} & 2016 & c/c++ & cf & 1412 & cfa \\
Costin et al.\,\cite{Costin2017Lua} & 2017 & lua & df & 1404, 1407, 1409, 1416 & dfa \\
EasyIVD\,\cite{Fang2017A} & 2017 & java & df & 1406, 1407 & pm, ps \\
OOPixy\,\cite{Nashaat2017Detecting} & 2017 & php & df & 1407, 1409, 1416 & dfa, ra, va \\
Scanner\,\cite{Chen2017Static} & 2017 & php & df & 1403, 1416, 1417 & dfa, se \\
SSD\,\cite{Obaida2017Interactive} & 2017 & java & df & 1403, 1416, 1417 & dfa \\
Stacy\,\cite{Lathar2017Stacystatic} & 2017 & c/c++ & cf & 1399, 1406, 1407, 1416 & cfa \\
SVAT\,\cite{Kalyanasundaram2017Static} & 2017 & c/c++ & cf & 1399, 1401, 1402, 1403, 1406, 1407, 1410, 1412, 1413, 1415, 1416, 1417 & se \\
TaintCrypt\,\cite{Rahaman2017Program} & 2017 & c/c++ & df & 1402, 1413 & dfa, se, smt \\
Tang et al.\,\cite{Tang2017Detecting} & 2017 & java & df & 1396, 1412, 1418 & dfa \\
Bianchi et al.\,\cite{Bianchi2018Broken} & 2018 & java & df & 1396 & dfa \\
Ferrara et al.\,\cite{Ferrara2018Tailoring} & 2018 & java & df & 1403, 1416, 1417 & dfa \\
Gao et al.\,\cite{Gao2018A} & 2018 & c/c++ & df, s & 1399, 1403, 1405, 1406, 1407, 1408, 1412, 1415, 1416 & ra, smt \\
HIP/SLEEK\,\cite{Bican2018Verification} & 2018 & c/c++ & df & 1399, 1406, 1407, 1412, 1415, 1416 & dfa \\
IFDS\,\cite{Belyaev2018Comparative} & 2018 & c\# & df & 1396, 1403, 1407, 1409, 1412, 1413, 1414, 1416, 1417, 1418 & dfa, se \\
IIFDroid\,\cite{Bohluli2018Detecting} & 2018 & java & df & 1403, 1416, 1417 & dfa \\
Panarotto et al.\,\cite{Panarotto2018Static} & 2018 & java & df & 1407, 1409 & dfa \\
Sails\,\cite{Cortesi2018Combining} & 2018 & java & df & 1403, 1416, 1417 & da, dfa, va \\
Zhou et al.\,\cite{Zhou2018Static} & 2018 & php & df & 1403, 1416, 1417 & se \\
Ares\,\cite{Li2019Ares} & 2019 & c/c++ & s & 1405 & pm \\
DCUAF\,\cite{Bai2019Effective} & 2019 & c/c++ & df & 1399, 1406, 1407, 1415, 1416 & cla \\
Debreach\,\cite{Paulsen2019Debreach} & 2019 & php & df & 1403, 1416, 1417 & cfa, dfa \\
DroidPatrol\,\cite{Talukder2019DroidPatrol} & 2019 & java & df & 1403, 1407, 1409, 1416, 1417 & cfa, dfa \\
Garmany et al.\,\cite{Garmany2019Static} & 2019 & c/c++ & df & 1416 & cfa, dfa, ra, se \\
IMSpec\,\cite{Gu2019IMSpec} & 2019 & c/c++ & df & 1396, 1399, 1405, 1406, 1407, 1408, 1412, 1415, 1416 & dfa, ra \\
Julia (Extension)\,\cite{Spoto2019Static} & 2019 & java & df & 1407, 1409, 1416 & dfa \\
LAID\,\cite{Xu2019A} & 2019 & c/c++ & df & 1399, 1406, 1407, 1408, 1416 & dfa, smt \\
Maskur et al.\,\cite{Maskur2019Static} & 2019 & php & df & 1407, 1409 & dfa \\
Nodest\,\cite{Nielsen2019Nodest} & 2019 & js & df & 1407, 1409 & dfa, pm \\
SSLDoc\,\cite{Gu2019SSLDoc} & 2019 & c/c++ & cf & 1396 & se \\
Taint-Things\,\cite{Schmeidl2019Security} & 2019 & jvm & df & 1403, 1416, 1417 & dfa \\
TSDroid\,\cite{Cao2019A} & 2019 & java & df & 1403, 1416, 1417 & dfa \\
UrFlow\,\cite{Chlipala2019Static} & 2019 & ur & cf & 1403, 1416, 1417 & se, smt \\
Vanguard\,\cite{Situ2019Automatic} & 2019 & c/c++ & df & 1406, 1407 & dfa \\
Ying et al.\,\cite{Ying2019Detecting} & 2019 & c/c++ & cf, df & 1399, 1406, 1407, 1416 & cfa, dfa \\
Ahmed et al.\,\cite{Ahmed2020Identifying} & 2020 & unspecified & s & 1416 & pm \\
ANTaint\,\cite{Wang2020Scaling} & 2020 & java & df & 1403, 1416, 1417 & dfa \\
AuthCheck\,\cite{Piskachev2020AuthCheck} & 2020 & java & cf & 1396 & ta \\
AVGuardian\,\cite{Hong2020AVGuardian} & 2020 & c/c++ & df & 1396 & dfa \\
CTAN\,\cite{Andarzian2020Compositional} & 2020 & c/c++ & df & 1403, 1416, 1417 & dfa, se \\
DBloop\,\cite{Luo2020Static} & 2020 & c/c++ & cf, df & 1399, 1406, 1407, 1416 & cfa, dfa, ps, ra \\
DepTaint\,\cite{Li2020DepTaint} & 2020 & c/c++ & df & 1399, 1403, 1406, 1407, 1408, 1412, 1415, 1416 & dfa \\
DroidRista\,\cite{Alzaidi2020DroidRista} & 2020 & java & df & 1403, 1416, 1417 & dfa \\
ELAID\,\cite{Xu2020ELAID} & 2020 & c/c++ & df & 1399, 1406, 1407, 1408, 1416 & dfa, smt, ta \\
ESLint\,\cite{Rafnsson2020Fixing} & 2020 & js & s & 1407, 1409 & pm \\
Graft\,\cite{Keirsgieter2020Graft} & 2020 & java & df & 1399, 1401, 1403, 1405, 1406, 1407, 1410, 1412, 1416 & dfa, ra \\
iDEA\,\cite{Bai2020IDEA} & 2020 & c/c++ & df & 1399, 1401, 1406, 1407, 1410, 1415, 1416 & cfa, dfa, se \\
Mandal et al.\,\cite{Mandal2020Crossprogram} & 2020 & java, c\# & df & 1403, 1416, 1417 & dfa \\
PGFit\,\cite{Nobakht2020PGFIT} & 2020 & java, jvm & cf & 1396, 1412, 1418 & cfa \\
Pogliani et al.\,\cite{Pogliani2020Detecting} & 2020 & robotic dsls & df & 1412 & dfa \\
SCFMSP\,\cite{Pouyanrad2020SCFMSP} & 2020 & c/c++ & df & 1403, 1416, 1417 & cfa, da, dfa \\
VulArcher\,\cite{Qin2020Vulnerability} & 2020 & java & df & 1396, 1403, 1407, 1409, 1416 & dfa, pm \\
AIT\,\cite{Yousaf2021Efficient} & 2021 & c/c++ & df & 1396, 1401, 1403, 1410, 1412, 1416, 1418 & va \\
Brinza et al.\,\cite{Brinza2021Virtual} & 2021 & java, php, py, js & df & 1403, 1406, 1407, 1409, 1416 & dfa, ra \\
CogniCrypyt\,\cite{Kruger2021CrySL} & 2021 & java & df & 1396 & dfa, va \\
Iqbal et al.\,\cite{Iqbal2021Enhancement} & 2021 & c/c++ & df & 1399, 1406, 1407, 1416 & dfa \\
K-MeLD\,\cite{Emamdoost2021Detecting} & 2021 & c/c++ & cf, df & 1416 & cfa, dfa \\
MEBS\,\cite{Zhang2021MEBS} & 2021 & c/c++ & df & 1399, 1405, 1406, 1407, 1412, 1415, 1416 & dfa, ra \\
MirChecker\,\cite{Li2021MirChecker} & 2021 & rust & df & 1399, 1406, 1407, 1408, 1416 & cfa, dfa, se, smt \\
ModGuard\,\cite{Dann2021Modguard} & 2021 & java & df & 1403, 1416 & dfa \\
Paramitha et al.\,\cite{Paramitha2021Static} & 2021 & php & df & 1404, 1406, 1407, 1409, 1416 & dfa \\
Rahaman et al.\,\cite{Rahaman2021Algebraicdatatype} & 2021 & java & df & 1403, 1416, 1417 & dfa \\
SADA\,\cite{Bai2021Static} & 2021 & c/c++ & df & 1399, 1406, 1407, 1416 & dfa, ra \\
SAND\,\cite{Lyu2021SAND} & 2021 & java & df & 1407, 1409 & cfa, dfa \\
SecuCheck\,\cite{Piskachev2021SecuCheck} & 2021 & java & df & 1406, 1407 & dfa, ra \\
TAJStaint\,\cite{Almashfi2021Static} & 2021 & js & df & 1406, 1407 & dfa \\
VulChecker\,\cite{Chen2021VulChecker} & 2021 & php & df & 1407, 1409 & cfa, dfa, pm \\
VulDetector\,\cite{Cui2021VulDetector} & 2021 & c/c++ & s & 1399, 1403, 1405, 1406, 1407, 1412, 1416, 1417 & pm, ps \\
FFIChecker\,\cite{Li2022Detecting} & 2022 & rust, c/c++ & cf, df & 1405, 1412, 1416 & cfa, dfa \\
IFLOW\,\cite{Yavuz2022Security} & 2022 & c/c++ & df & 1396 & dfa \\
IncreLux\,\cite{Zhai2022Progressive} & 2022 & c/c++ & df & 1399, 1406, 1407, 1416 & dfa, pm, ra, se \\
MAD\,\cite{Wei2022An} & 2022 & c/c++ & df & 1399, 1406, 1407, 1412, 1415, 1416 & dfa, ra \\
MPChecker\,\cite{Lu2022Detecting} & 2022 & java & df & 1396 & dfa \\
MVDetecter\,\cite{Nie2022MVDetecter} & 2022 & c/c++ & cf, df & 1399, 1405, 1406, 1407, 1408, 1412, 1415, 1416 & dfa, pm \\
NDI\,\cite{Zhou2022NonDistinguishable} & 2022 & c/c++ & cf & 1399, 1405, 1406, 1407, 1412, 1416 & ra, se \\
PATA\,\cite{Li2022PathSensitive} & 2022 & c/c++ & cf & 1405, 1412, 1416 & ra, smt, ta \\
Peahen\,\cite{Cai2022Peahen} & 2022 & c/c++ & cf & 1401, 1410, 1416 & ra \\
PSCAT\,\cite{Ma2022Code} & 2022 & py & cf, df & 1396, 1403, 1407, 1409, 1415, 1416, 1417 & cfa, dfa \\
Sausage\,\cite{Elgharabawy2022SAUSAGE} & 2022 & java & df & 1396, 1399, 1406, 1407, 1416 & dfa \\
Seader\,\cite{Zhang2022ExampleBased} & 2022 & java & s & 1396 & pm \\
StaticFac\,\cite{Lu2022Static} & 2022 & c/c++ & cf & 1396 & cfa, se \\
TaintCrypt\,\cite{Rahaman2022From} & 2022 & c/c++ & df & 1402, 1413 & dfa, se \\
Tracer\,\cite{Kang2022TRACER} & 2022 & c/c++ & df & 1399, 1401, 1403, 1406, 1407, 1408, 1409, 1410, 1412, 1415, 1416 & dfa, pm \\
Wasmati\,\cite{Brito2022Wasmati} & 2022 & c/c++ & df & 1399, 1406, 1407, 1408, 1412, 1415, 1416 & dfa \\
Zhuang et al.\,\cite{Zhuang2022A} & 2022 & c/c++ & cf & 1401, 1410 & dfa, se \\
Alqaradaghi et al.\,\cite{Alqaradaghi2023Design} & 2023 & java & s & 1413 & pm \\
Aria\,\cite{Abolhassani2023A} & 2023 & java & cf, df & 1403, 1416, 1417 & la \\
CompTaint\,\cite{Banerjee2023Compositional} & 2023 & java & df & 1396, 1398, 1404, 1406, 1407, 1409, 1415, 1416 & dfa, ra, ta, va \\
FLuaScan\,\cite{Li2023Finding} & 2023 & lua & df & 1406, 1407 & dfa \\
HM-SAF\,\cite{Zhu2023HMSAF} & 2023 & java & df & 1403, 1416, 1417 & cfa, dfa \\
ISDE\,\cite{An2023Refining} & 2023 & c/c++ & df & 1399, 1406, 1407, 1415, 1416 & dfa \\
Lockpick\,\cite{Cai2023Place} & 2023 & c/c++ & df & 1401, 1410, 1416 & cla, ra, ta \\
Opdebeeck et al.\,\cite{Opdebeeck2023Control} & 2023 & ansible & cf, df & 1396, 1402, 1403, 1411, 1412, 1413, 1414, 1416, 1418 & cfa, dfa \\
PeX\,\cite{Zhou2023Automatic} & 2023 & c/c++ & df & 1396 & cfa, dfa, ra \\
PrivDroid\,\cite{El2023PrivDroid} & 2023 & java & df & 1396 & dfa \\
rat\,\cite{Parolini2023Sound} & 2023 & py & s & 1416 & pm \\
RedSoundRSE\,\cite{Tiraboschi2023Sound} & 2023 & unspecified & cf & 1403, 1416, 1417 & da, se, smt \\
RVDetecor\,\cite{Tao2023Vulnerability} & 2023 & c/c++ & df & 1399, 1406, 1407, 1416 & dfa \\
SAMVA\,\cite{Gicquel2023SAMVA} & 2023 & c/c++ & cf & 1407, 1409 & pm, ta \\
TaintMini\,\cite{Wang2023Taintmini} & 2023 & java & df & 1403, 1416, 1417 & dfa \\
TaintSA\,\cite{Tang2023Android} & 2023 & java & df & 1403, 1416, 1417 & dfa, ra \\
UACatcher\,\cite{Ma2023When} & 2023 & c/c++ & cf & 1399, 1406, 1407, 1415, 1416 & cfa, ra \\
VulPathsFinder\,\cite{Zhao2023VulPathsFinder} & 2023 & php & df & 1396, 1407, 1409, 1416 & dfa, ra \\
Wang et al.\,\cite{Wang2023Detecting} & 2023 & java & df & 1403, 1416, 1417 & dfa, ra \\
Zhang et al.\,\cite{Zhang2023Research} & 2023 & java & df & 1403, 1407, 1409, 1416, 1417 & ra \\
BinSweep\,\cite{Oldani2024Binsweep} & 2024 & c/c++ & cf & 1416 & cfa \\
BolaRay\,\cite{Huang2024Detecting} & 2024 & php & cf, df & 1396, 1406, 1407, 1412, 1413, 1418 & cfa, dfa \\
Comchecker\,\cite{Ding2024Security} & 2024 & py, java & s & 1412, 1416, 1418 & la \\
Cryptolation\,\cite{Frantz2024Methods} & 2024 & py & cf & 1396 & cfa, dfa, la, ps \\
CtChecker\,\cite{Zhou2024CtChecker} & 2024 & c/c++ & df & 1403, 1416, 1417 & dfa, ra \\
CtyptoPyt\,\cite{Guo2024CryptoPyt} & 2024 & py & df & 1396, 1402, 1403, 1412, 1413, 1414, 1416, 1417, 1418 & dfa \\
Deng et al.\,\cite{Deng2024Static} & 2024 & jvm & df & 1403, 1416, 1417 & dfa \\
Díez-Franco et al.\,\cite{Díez-Franco2024Optimized} & 2024 & c/c++ & cf, df & 1396, 1403, 1416 & dfa, ra \\
EffiTaint\,\cite{Li2024EffiTaint} & 2024 & java & df & 1403, 1416, 1417 & dfa, ra \\
ERASan\,\cite{Min2024ERASan} & 2024 & rust & df & 1399, 1406, 1407, 1416 & ra, ta \\
FuncTion-V\,\cite{Remil2024Automatic} & 2024 & c/c++ & cf & 1406, 1407 & se \\
Gopher\,\cite{Zhang2024Gopher} & 2024 & go & df & 1396 & dfa, pm, ps \\
Graph.js\,\cite{Ferreira2024Efficient} & 2024 & js & df & 1404, 1406, 1407, 1409, 1415, 1416 & dfa \\
JWTKey\,\cite{Xu2024JWTKey} & 2024 & java & df & 1396, 1403, 1416 & dfa, ps \\
Kluban et al.\,\cite{Kluban2024On} & 2024 & js & df & 1404, 1407, 1409, 1415, 1416 & dfa \\
LibAlchemy\,\cite{Wu2024LIBALCHEMY} & 2024 & c/c++ & cf, df & 1399, 1405, 1406, 1407, 1412, 1415, 1416 & cfa, da, ra, smt \\
MangoDFA\,\cite{Gibbs2024Operation} & 2024 & c/c++ & df & 1399, 1406, 1407, 1409, 1416 & cfa, da, dfa \\
MiniBLE\,\cite{Zhang2024MiniBLE} & 2024 & js & s & 1396 & pm \\
MiniCat\,\cite{Zhang2024MiniCAT} & 2024 & js & df & 1411, 1413 & dfa \\
MQRRactic\,\cite{Yuan2024MQTTactic} & 2024 & erlang, java, js & cf & none & cfa, mc, ra, se \\
MtdScout\,\cite{Zhang2024MtdScout} & 2024 & java & s & 1412, 1418 & pm \\
OctupusTaint\,\cite{Qasem2024OctopusTaint} & 2024 & c/c++ & df & 1406, 1407 & cfa, dfa \\
Partyka et al.\,\cite{Partyka2024Hardcoded} & 2024 & java, jvm & s & 1396, 1412, 1413, 1414, 1418 & va \\
Pinjari et al.\,\cite{Pinjari2024Integrity} & 2024 & java & s & 1411, 1413 & pm \\
RbSFP\,\cite{Gershfeld2024Evaluating} & 2024 & c/c++ & cf, df & 1399, 1402, 1405, 1406, 1407, 1408, 1410, 1412, 1413, 1416 & pm \\
Samba\,\cite{Liu2024Samba} & 2024 & c/c++ & df & 1396 & dfa, pm \\
SaTC\,\cite{Chen2024SaTC} & 2024 & java & df & 1396, 1399, 1406, 1407, 1409, 1416 & dfa \\
Semgrep\,\cite{Kree2024Using} & 2024 & php & df & 1396, 1402, 1407, 1409, 1413, 1415, 1416 & dfa, pm \\
SerdeSniffer\,\cite{Liu2024SerdeSniffer} & 2024 & java & df & 1415, 1416 & dfa, pm, ra \\
ST-Checker\,\cite{Zhao2024Static} & 2024 & java & cf, df & 1402, 1412, 1413, 1416 & dfa, se, smt \\
STASE\,\cite{Shafiuzzaman2024STASE} & 2024 & c/c++ & df & 1396, 1399, 1406, 1407, 1408, 1415, 1416 & dfa, pm, ps, se, smt \\
Teodorescu et al.\,\cite{Teodorescu2024Static} & 2024 & c/c++ & df & 1399, 1406, 1407, 1415, 1416 & cfa, dfa \\
UQuery\,\cite{Huang2024UQuery} & 2024 & php & df & 1401, 1407, 1409, 1410, 1411, 1413, 1416 & dfa, se \\
VAScanner\,\cite{Zhang2024Does} & 2024 & java & df & 1403, 1404, 1412, 1413, 1414, 1415, 1416, 1417, 1418 & dfa, ra \\
VulnSlicer\,\cite{Liao2024VulnSlicer} & 2024 & java & df & 1396, 1403, 1407, 1409, 1415, 1416 & dfa, pm, ps \\
WasmChecker\,\cite{Zhuang2024WasmChecker} & 2024 & c/c++ & cf & 1399, 1406, 1407, 1408, 1412, 1415, 1416 & se, smt \\
WFinder\,\cite{Ma2024A} & 2024 & c/c++ & df & 1399, 1403, 1406, 1407, 1409, 1416 & dfa \\
Yuga\,\cite{Nitin2024Yuga} & 2024 & rust & cf & 1399, 1403, 1406, 1407, 1415, 1416 & pm, ra
\label{tab:analyzers}
\end{tabularx}
\normalsize
\vspace{-.1cm}

%% file: 05_Discussion.tex
\section{Discussion}

Our results reveal the current landscape of static security analysis research. 
Based on these findings, we outline key challenges and future directions that require action from the research community.

\subsection{Lifting Analysis to Fewer Vulnerabilities of Higher Severity}
Our study shows that only a minority of static security analyzers report detecting exploitable vulnerabilities (11.8\%) or confirmed vulnerabilities (18.3\%).
Examining the types of weaknesses these analyzers target reveals that most findings are low-level, such as potential buffer overflows. 
For example, the narrow CWE-119 (buffer overflow) is the third most commonly addressed weakness, with 40 analyzers covering it.
While such findings are important, they need to be interpreted in the context of the entire system to assess their real-world impact\,\cite{Peldszus2026}.
Due to various reasons, some vulnerabilities are harder to exploit than others\,\cite{Nayak2014}.
For a buffer overflow to be truly exploitable, an attacker must be able to propagate malicious input to the vulnerable code location, a condition that is typically outside the scope of standard overflow checks. 

One way to address this limitation is by combining low-level analyses, such as integrating buffer overflow detection with taint analysis. 
Analysis coupling is an emerging approach that moves beyond isolated low-level findings towards more meaningful, system-level insights\,\cite{Reiche2025,Reiche2025a}. 
Preliminary work demonstrates that integrating static code analysis results into architectural models improves the detection of vulnerabilities that might otherwise be overlooked. 
Moreover, the formalization of coupling conditions provides a systematic framework for combining analyses operating at different abstraction levels, enhancing security assurance. 
Similarly, combining multiple analyses at the implementation level can yield findings that are more actionable and context-aware.

Finally, our review highlights a general lack of attention to result presentation. 
Developers, however, require clear and actionable outputs to effectively remediate vulnerabilities\,\cite{vassallo2020developers}. 
Integrating coupled analyses with thoughtful result presentation, such as highlighting critical paths, contextualizing low-level alerts, and summarizing systemic risks, can improve adoption and practical impact. 
At the same time, prioritization based on coupled analyses will lead to fewer, more meaningful findings, therefore, addressing the overwhelming number of findings that developers face\,\cite{Walden2014}.

\subsection{Covering All Aspects of Software Security}
Another notable trend is the strong focus on a small set of well-known vulnerabilities.
Many analyzers target issues that frequently appear in the OWASP Top 10, such as injection flaws and memory safety problems.
However, other issues, like insecure design and security misconfigurations, receive comparatively little attention, although they also appear in the OWASP Top 10.
Albeit important, this focus suggests that static security analysis research often gravitates toward the “low-hanging fruit.”
Similarly, only a minority of static security analyzers explicitly target security features such as access control, although their practical relevance is known for a long time\,\cite{Dalton2009}.

Future work should place stronger emphasis on understudied vulnerabilities such as insecure design, configuration errors, and supply chain risks, which have been added to the OWASP Top 10 in 2025.
Investigating whether and how static analysis techniques can be extended to cover these vulnerabilities is an important open research gap.

\subsection{Increase Relevance by Customization of Analyzers}
As our results show, most static security analyzers are designed to be domain-independent, aiming for broad applicability across domains.
However, this introduces a tradeoff between coverage and specificity.
As they do not capture the domain-specific context\,\cite{Pina2021,Peldszus2026}, such analyzers may miss or underprioritize vulnerabilities that are relevant in particular projects.
In contrast, domain-specific analyzers can provide deeper and more accurate checks, but are far less researched.

Furthermore, we found that most analyzers rely on hardcoded check specifications that cannot be adapted to project-specific needs.
While many tools aim for a proof of concept approach, it limits adaptability and makes it harder for practitioners to transfer analyzers to specific domains, or projects\,\cite{nunes2019}.
Static security analyzers should become more adaptable by allowing users to define custom checks and to configure analyses for specific domains, or scenarios.

\subsection{Insufficient Evaluation and Reporting of Limitations}
The observed lack of systematic evaluation and transparent reporting of limitations remains a central challenge in static security analysis research.
Most analyzers are assessed using custom-built benchmarks, which are typically small in scale (often fewer than 20 systems).
Such benchmarks are susceptible to selection bias and hinder meaningful comparison across analyzers, limiting the generalizability of results. 
Quality standards for benchmarking\,\cite{acm_standards} are not always met.
In addition, the absence of reliable ground truth for real-world systems makes it difficult to compute robust statistical measures such as precision and recall, although negative test cases are a particularly relevant element\,\cite{Miltenberger2023}. 
This can undermine developer trust in the tools. 
One reason could be that existing standardized benchmarks are limited to specific programming languages or vulnerability classes, reducing their broader applicability. 
The community would benefit from more standardized, large-scale benchmarks that span multiple languages and vulnerability types. 

We further observed that attacker models are rarely made explicit, especially in papers published outside dedicated security venues.
The absence of clearly defined attacker capabilities and assumptions makes it difficult to assess whether reported results are realistic or practically relevant, potentially leading to unrealistic security claims.
The most frequently reported technical limitations concern over- and under-approximations.
Another recurring limitation is the restricted support for dynamic language features.
Although some analyzers do handle selected dynamic constructs (e.g., for Python or JavaScript), many tools draw the line when faced with reflection, dynamic code loading, or runtime code generation.
This limits their applicability to real-world software systems that heavily rely on such features, and leads to many false negatives and false positives.
We call for more systematic reporting of limitations, assumptions, and attacker models.
Clear documentation of the limitations of static security analyzers, under which threat assumptions, would improve transparency, and make research results more actionable for practitioners.

%% file: 07_Conclusion.tex
\section{Conclusion}
We conducted a large-scale systematic literature review covering 246 static security analyzers.
We examined the code elements analyzed, detectable vulnerabilities, analysis techniques, evaluation practices, and reported limitations. 
Most analyzers are domain-independent, focus on languages such as Java and C, and dataflow analysis on control flow graphs is the predominant technique.
Checks repeatedly target well-known issues like SQL injection, cross-site scripting, and buffer overflows, but analyzers mainly detect low-level coding vulnerabilities of which only a minority has been proven to be exploitable.
Only 11.8\% of the analyzers claim to have detected at least one exploitable vulnerability.
Since nearly all static security analyzers are evaluated with custom benchmarks, the lack of a ground truth for real-world systems makes it difficult to evaluate statistical metrics for static security analyzers.
More than half of the papers do not systematically report limitations of their static security analyzers.

Overall, static security analysis would benefit from standardized, widely applicable benchmarks, broader coverage of understudied vulnerability classes (e.g., insecure design, misconfigurations, logging weaknesses, and supply chain risks), and stronger emphasis on the practical relevance of detected issues. 
Combining multiple low-level findings to lift analysis results to the scope of the system could increase the relevance of findings.
More systematic reporting of assumptions, attacker models, and limitations is essential to improve transparency and real-world applicability.